\documentclass{aa}
\usepackage[varg]{txfonts}
\usepackage{graphicx}
\usepackage{psfig}      
\usepackage{color}
\usepackage{txfonts}    

\begin{document}

\title{Extent and structure of intervening absorbers from absorption lines  redshifted 
on quasar emission lines\thanks{Based on observations with VLT-UVES and Keck-HIRES} } 
\titlerunning{Structure of intervening absorbers} 

\author{J. Bergeron\inst{1} and P. Boiss\'e\inst{1} }
\institute{$^{1}$ Institut d'Astrophysique de Paris, UPMC - CNRS,  UMR 7095, 98 bis Bd Arago,  
F-75014, Paris, France  \\
       \email{bergeron@iap.fr} }
           
\date{Received date 31/03/2017/ 
Accepted date }

\abstract
{}
{ We wish to study the extent and sub-parsec spatial structure of intervening 
quasar absorbers, mainly those involving cold neutral and molecular gas.} 
{ We have selected quasar absorption systems with high spectral resolution and a good 
signal-to-noise ratio data, with some of their lines falling on quasar emission 
features. 
By investigating the consistency of absorption profiles seen for 
lines formed either against the quasar continuum source or on the much more 
extended (Ly$\alpha$-N\,{\sc v}, C\,{\sc iv} or Ly$\beta$-O\,{\sc vi}) emission 
line region (ELR), we can probe the extent and structure of the foreground absorber
over the extent of the ELR ($\sim 0.3-1$ pc). 
The spatial covering analysis provides constraints on the transverse size of the absorber and 
thus  is complementary to variability or photoionisation modelling studies, 
which yield information on the absorber size along the line of sight. 
The methods we used to identify spatial covering or structure effects involve line profile 
fitting and curve-of-growth analysis.}
{ We have detected three absorbers with unambiguous non-uniformity effects in neutral gas. 
For the extreme case of the Fe\,{\sc i} absorber at $z_{\rm abs}=0.45206$ towards 
HE~0001$-$2340, we derive a coverage factor of the ELR of at most 0.10 and possibly 
very close to zero; this implies an overall absorber size no larger than 0.06 pc. 
For the $z_{\rm abs}=2.41837$ C\,{\sc i} absorber towards QSO J1439$+$1117, 
absorption is significantly stronger towards the ELR than towards 
the continuum source in several C\,{\sc i} and C\,{\sc i}$^{\star}$ velocity components, pointing 
to spatial variations of their column densities of about a factor of two and to structures 
at the 100 au - 0.1 pc scale. The other systems with firm or possible
effects can be described in terms of a partial covering of the ELR, with 
coverage factors in the range 0.7 - 1. The overall results for cold neutral 
absorbers imply a transverse extent of about five times the ELR size or smaller, 
which is consistent with other known constraints. 
Although not our primary goal, we also checked when possible that singly-ionised
absorbers are uniform at the parsec scale, in agreement with previous studies. In 
Tol 0453$-$423, we have discovered a very unusual case with a small but clearly significant 
residual flux for a saturated Fe\,{\sc ii}$\lambda2600$ line at $z_{\rm abs}=0.72604$
seen on Ly$\alpha$ emission, thus with an absorber size comparable 
to or larger than that  of the ELR. 
}
{}


\maketitle

\section{Introduction}

Some of the absorption systems detected in distant quasar spectra that are commonly used to investigate 
the properties of diffuse gas in various environments throughout the universe have 
been noted to display unexpected properties in the relative strength of absorption lines that are associated 
with distinct transitions from a given species. For instance, in some systems with broad absorption lines (BAL) or in
intrinsic systems, the observed resolved profiles of multiplet lines are clearly inconsistent 
with models in which the absorber uniformly covers the background source (Barlow \& Sargent
1997; Ganguly et al. 1999). This is especially 
evident when resolved line profiles display a flat core, which indicates saturation together with a 
non-zero residual flux at the line centre. In BAL systems, such effects can be 
easily explained by invoking small cloudlets within the disc wind that only 
partially cover the quasar accretion disc. 

A variant of this situation is found when the profile of  intervening  
absorption lines that fall on 
top of quasar emission features appears to be inconsistent with the profile of lines from the same species that formed 
against the continuum source alone. One good example is the C\,{\sc i} and H$_2$ 
system at $z_{abs} = 2.3377$ towards LBQS 1232+082 (Balashev et al. 2011). 
The C\,{\sc i}$\lambda 1656$ feature arises on the C\,{\sc iv} quasar emission line; thus, the 
background flux at the corresponding wavelength is provided by both the accretion 
disc that is responsible for the continuum emission and by the C\,{\sc iv} 
emission line region (hereafter ELR), while the C\,{\sc i}$\lambda 1560$  line is seen against the 
continuum source alone. While the accretion disc is no larger than about 100 au 
(Dai et al. 2010), the ELR is much more extended, with a size in the range 0.1 - 1 pc for luminous 
quasars (Bentz et al. 2009). Since the C\,{\sc i}$\lambda 1656$ feature 
is weaker than C\,{\sc i}$\lambda 1560$, whereas the opposite is expected on the basis of 
oscillator strength values, one cannot escape the conclusion that the C\,{\sc i} absorber is not
uniform over the whole background source and covers the extended ELR only partially. 
A few other systems that display effects of this type have been reported (see, e.g., Krogager et al. 
2016; Fathivavsari et al. 2017); 
we note, however, that the interpretation for some of them remains ambiguous because the presence of 
unresolved optically thick velocity components might be sufficient to explain the observed line ratio. 

When present, these effects imply a lower apparent opacity for the lines that are affected by partial covering. This 
biases the determination of column densities, $N$, and Doppler parameters, $b$. Thus, an immediate 
objective is to properly take these effects into account in order to obtain correct $N$ and $b$ values. 
A further more essential motivation is related to our knowledge of the size and internal structure 
of the associated gaseous clouds. In the intervening systems mentioned above, the peculiar 
relative strength of absorption lines is related to the finite extent of the 
background source and more specifically to its composite nature, involving two widely different 
scale lengths ($\approx$ 100~au and 1~pc).  By modelling the profile of absorption 
lines that fall on or away from quasar emission features, one should then be able to compare the 
absorber size to these scale lengths. 

This approach is especially relevant for molecular or neutral gas (as probed, e.g., by C\,{\sc i} 
lines for high-redshift systems) 
since cloud sizes in the range 0.1 - 10 pc have been 
inferred (see Jenkins \& Tripp 2011 for Galactic gas and Jorgenson et al. 2010 for high-redshift 
systems), which is comparable to the ELR size. These scales have been derived from the analysis of 
C\,{\sc i} fine-structure transitions, which provides an estimate of the volume density; the inferred 
extent is therefore  along the line of sight. In contrast, the analysis of partial
covering effects yields constraints on the  transverse size. Thus, both methods provide 
independent complementary estimates, and their comparison should lead to a robust estimate of the 
absorber extent, which is a key parameter for modelling.

The absorption lines induced by an absorber located in front of an extended background source is
governed not only by their relative extent, but also by the small-scale structure within 
the intervening gas. In the interstellar medium of our own Galaxy, small-scale structure in the neutral
medium (as traced by C\,{\sc i}  or Na\,{\sc i}) is observed at all scales above about 10 au 
(Welty 2007; Watson \& Meyer 1996). If structure over such small scales is also present in high-redshift 
galaxies, it should manifest itself through time changes in quasar absorption lines, as 
argued recently by Boiss\'e et al. (2015). Transverse peculiar velocities of a few 100 
km s$^{-1}$ are expected for the observer,
quasars, and intervening galaxies. This implies drifts of the line of sight through the
absorber of tens to hundreds of au over a time interval of about 10 years. 
To date, only tentative variations (3$\sigma$ significance level) have been observed for neutral 
(C\,{\sc i}) and  molecular (H$_2$) gas in a damped Ly$\alpha$ (DLA) absorber at $z_{\rm abs}=2.05454$ 
towards FBQS J2340$-$0053 over a two-year time interval 
(Boiss\'e et al. 2015, and references therein for other approaches involving quasar pairs or lensed 
quasars). If internal structure were present in the intermediate 100 au - 1pc range within distant
neutral or molecular absorbers, this would potentially affect the behaviour of absorption lines 
that are detected on quasar emission features. Thus, a detailed analysis of such 
absorption lines can provide useful information that complements the information provided by time 
variation studies.

While these effects have been investigated in many quasar intrinsic systems (see, e.g., Hamann et 
al. 2011), to our knowledge no systematic study has been performed for  intervening 
systems. Only a few cases have been identified, suggesting that these effects are rare, while given 
the similarity between the size inferred for the C\,{\sc i} absorbers and the ELR extent (Jorgenson 
et al. 2010), one might 
expect them to be common. In order to clarify this question and derive useful constraints on the 
extent and small-scale structure of distant neutral absorbers, we have assembled a sample of 
C\,{\sc i}, Fe\,{\sc i,} and H$_2$ systems detected in quasar spectra with a good signal-to-noise ratio (S/N) and high 
resolution,  and we investigate their properties in a systematic manner.

This paper is organised as follows. In Sect. 2 we describe the various effects 
that can be expected when an absorber is not uniform over the whole extent of the ELR together with 
the simplest models that can be used to account for them. We also discuss under which conditions 
the non-uniformity can be established unambiguously. Section 3 presents the sample we studied and 
the analysis we performed in order to search for partial covering or structure effects. 
Our results on the transverse extent of the absorbers are presented in Sect. 4.
A discussion together with future prospects are given in Sect. 5.

\section{Effects due to non-uniform absorbers}

\subsection{Expected effects}
When an absorber is not uniform over the extent of the background source, two alternatives 
are possible.  
First, the absorption towards the ELR (that is, seen in a quasar emission line) can be  
 weaker than what is expected on the basis of absorption lines that are detected against the continuum 
source alone. This corresponds to the classical partial covering effect (Fig. 1ab). 
In the most extreme situation, the fraction of the ELR that iscovered can be so small that the emission 
line flux remains essentially unabsorbed (we recall that the size of the ELR is at least two orders 
of magnitude larger than that of the accretion disc). 
   \begin{figure}
\hspace{7mm}
        \includegraphics[width=8.cm,angle=0]{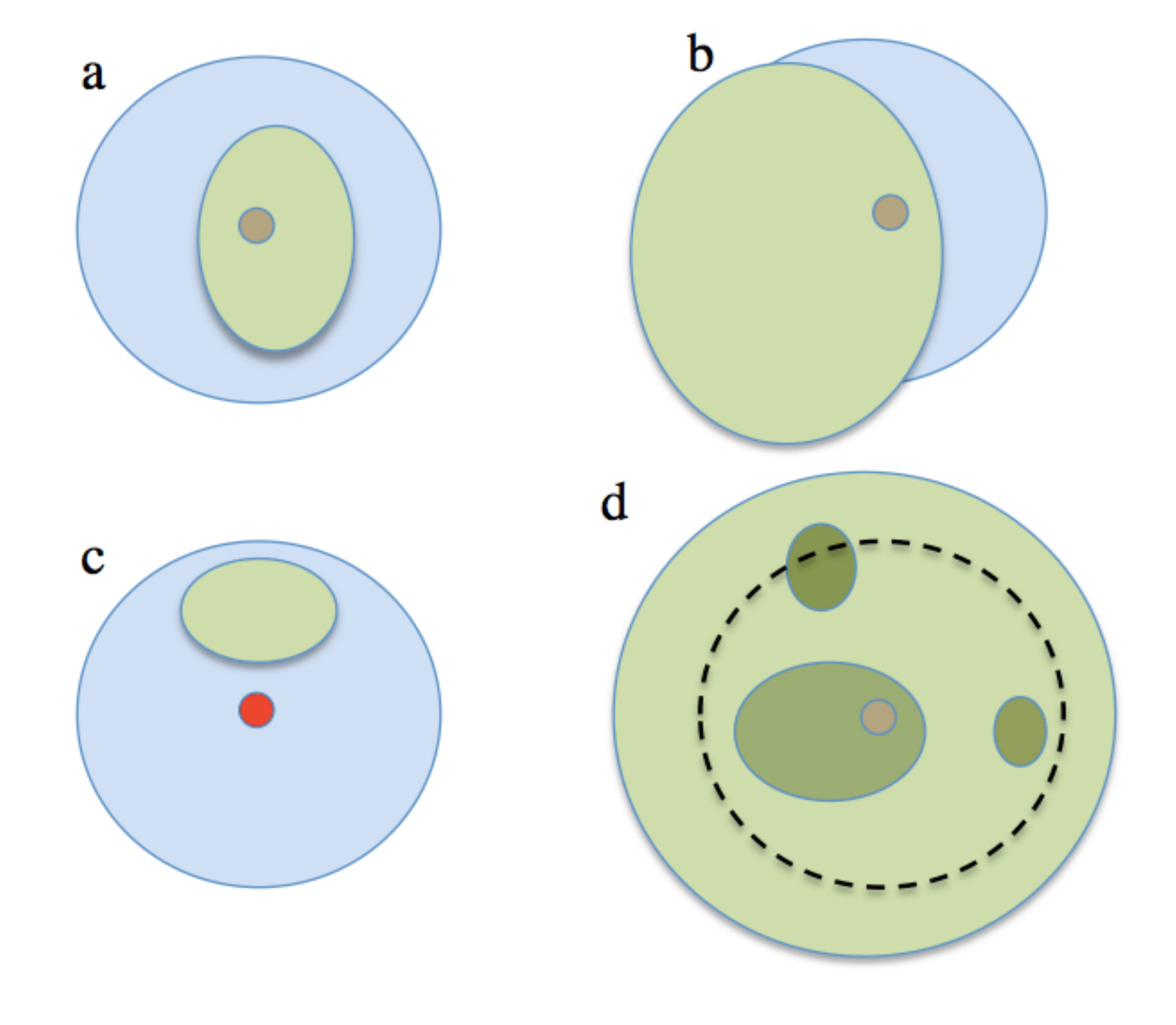}
    \caption{Four configurations for the quasar light source and foreground absorber. The continuum 
source (i.e., accretion disc) appears in red 
(or black when occulted), while the extended emission line region (ELR) is
plotted in light blue. The absorber (green) may cover the continuum source, but only part
of the ELR (a,b), as assumed in the partial covering model. It may also
cover part of the ELR, but not at all the continuum source (c), or, more likely,
cover the whole quasar source but in a non-uniform way, if internal
structure is present (yellow areas correspond to regions inducing weaker absorption; 
the black dotted line delineates the ELR). In cases a and b, absorption is stronger 
towards the continuum source than towards the ELR, whereas 
it is weaker in cases c and d. }       
   \label{PB}   
   \end{figure} 
\\
Second, the absorption seen against the ELR can be  stronger than  
that detected on the continuum; in Figs. 1c and 1d, 
we show two configurations corresponding to this alternative. The absorber in Fig. 1c 
does not absorb the continuum source flux at all and that in Fig. 1d is inhomogeneous, with an 
average absorption of the emission line flux larger than that of the continuum flux. 

\subsection{Models}
\subsubsection{Partial covering model}
Several authors have proposed models that can be used to describe absorption by a 
non-uniform gas layer. The most popular is the partial covering model (Barlow \& Sargent 
1997; Hamann et al. 1997, Ganguly et al. 1999), in which the absorber is assumed to 
uniformly cover some fraction of the source with an opacity $\tau(\lambda)$, the remaining 
fraction being unabsorbed. In this picture, the observed flux writes 
\begin{equation}
F_{obs}(\lambda) = C_f F_b \exp(-\tau(\lambda)) + (1 - C_f) F_b, 
\end{equation}
where $F_b$ is the total background flux ($F_b = F_c + F_{elr}$, with $F_c$ and $F_{elr}$ 
referring to the continuum and ELR sources, respectively), $\tau$ is the line opacity and $C_f$ 
the covering factor, that is, the fraction of the background flux that is covered by the absorber. 
In this equation, $F_b$ varies smoothly with wavelength 
over the interval covered by the absorption line. The term $\exp(-\tau(\lambda))$ describes 
the absorption line profile. The corresponding normalised spectrum is 
\begin{equation}
F_{obs,n}(\lambda) = C_f  \exp(-\tau(\lambda)) + (1 - C_f).
\end{equation}
An immediate consequence of partial covering is that saturated lines do 
not reach the zero level, but display a non-zero line flux residual 
(LFR, after Balashev et al.  2011) at their core. For $\exp(-\tau)\ll1$, the above relation
gives 
\begin{equation}
F_{obs,n}(\mathrm{core})=\mathrm{LFR} = 1 - C_f.
\end{equation}
For resolved doublet lines of this type, the apparent opacity ratio is no longer equal 
to the atomic physics value, but reaches unity when both lines are optically thick and 
affected by the same $C_f$ value. The LFR can be read directly on the observed spectrum, 
providing $C_f$ ($C_f = 1 - \mathrm{LFR}$). If lines are 
not resolved, profile fitting or curve-of-growth analysis of at least two transitions
must be used the derive the LFR and $C_f$ values.

$C_f$ is determined by the fraction of the ELR and of the continuum flux that is occulted, 
$C_{elr}$ and $C_c$ , respectively, and by the ratio of the emission line to the continuum flux, 
$x =  F_{elr}/F_c$, at the wavelength of the absorption feature considered (we follow the 
notations introduced by Ganguly et al. (1999), except for $W$, which is commonly used for equivalent 
widths, for which we adopt $x$). From the relation $C_f F_b = C_{elr} F_{elr} + C_c F_c$, 
we obtain 
\begin{equation}
C_f  = \frac{C_c +x\, C_{elr}}{1+x} \ \ \ \mathrm{or, equivalently,} \ \ 
C_{elr}  = \frac{C_f\, (1+x) -C_c}{x}.  
\end{equation}
The configurations shown in Fig. 1ab correspond to $C_c = 1$ and $C_{elr} < 1,$ while 
$C_c = 0$ and $C_{elr} < 1$ correspond to the configuration sketched in Fig. 1c. 
In practise, the $x$ value 
can be estimated by interpolating the continuum measured on the blue and red side of the 
emission feature at the location of the absorption line (this is performed more 
accurately if a 
flux-calibrated spectrum is available). We stress that in this model the absorber is 
uniform (with the advantage of introducing a mininum number of free parameters) but the 
covering of the source is not. 
\subsubsection{Two-value model} 
The above picture would not be appropriate to describe absorbers in which the opacity 
towards the continuum source is lower than the value averaged over the entire 
background source, as in Fig. 1d.
In this case, one could instead consider a two-value model involving distinct opacities for 
the gas in front of the continuum source, $\tau_c(\lambda)$, and in front of the ELR, 
$\tau_{elr}(\lambda)$ (note that Ganguly et al. (1999) considered a sort of mixed model in their 
Appendix A2, involving two opacity values $\tau_{elr}$ and $\tau_{c}$ together with two covering 
factors $C_{c}$ and $C_{elr}$). With the notations introduced above, the observed profile writes
\begin{equation}
F_{obs}(\lambda) = F_c \exp(-\tau_c(\lambda)) +  F_{elr} \exp(-\tau_{elr}(\lambda)), 
\end{equation}
corresponding to the normalised profile, 
\begin{equation}
F_{obs,n}(\lambda) = \frac{\exp(-\tau_c(\lambda)) + x \exp(-\tau_{elr}(\lambda))}{1+x}.
\end{equation}

This model, with only two discrete opacity values, $\tau_{c}$ in front of the continuum 
source and $\tau_{elr}$ elsewhere may look academic (we note that the partial covering model
also assumes two discrete values, $\tau$ and 0, but their spatial distributions are different
in the two models). 
A more realistic picture is that of an absorber displaying spatial fluctuations of the opacity 
that are due to internal structure within the region probed by the ELR. If the fluctuations 
remain moderate enough, one may adequately represent the inhomogeneous gas layer by the two
parameters $\tau_{c}$ and $\tau_{elr}$, where $\tau_{elr}$ characterises 
the ``effective'' (i.e., ``equivalent uniform'') absorber intercepted by the ELR. 
If in this picture the continuum source is located behind a region of low 
(respectively large)  
column density, this will result in $\tau_{elr} > \tau_{c}$ 
(respectively $\tau_{elr} < \tau_{c}$), 
and the $\tau_{elr} / \tau_{c}$ ratio can be used as a measure of the deviation from a uniform 
covering.  This two-value model is more flexible than the partial covering picture and 
can potentially describe a wider range of physical situations. 
Since the underlying assumptions about the geometry are distinct in these types of models, 
there is no rigorous correspondence between them. In the optically thin limit,
however, one can obtain the $\tau_{c}$ and $\tau_{elr}$ values corresponding to a given 
partial covering model (characterised by $\tau$ and 
$C_f$) by using Eqs. (2) and (5) and by setting $C_c = 1$ 
as well as $\tau_{c} = \tau$ (Eqs. (2) and (5), which do not have the same functional form in 
general, become equivalent in the $\tau \ll1$ limit) . As expected, we obtain
$\tau_{elr} = C_{elr}\, \tau$, which is the average opacity value over the ELR extent.

If appropriate absorption lines formed against the continuum source can be used to derive 
$\exp(-\tau_c(\lambda)),$ then in the frame of this model, Eq. (6) provides 
the absorption profile that one would obtain against the ELR alone,
\begin{equation}
\exp(-\tau_{elr}(\lambda)) = \frac{(1+x) F_{obs,n} -\exp(-\tau_c(\lambda))}{x}, 
\end{equation}
which potentially allows us to derive $\tau_{elr}$ and next to compare it to $\tau_c$. 
We note that the above relations are still valid if the convolution by the instrumental line 
spread function is taken into account. 


We finally mention that some other non-uniform models such as the power-law model have 
been introduced by Arav et al. (2005) in the context of intrinsic absorbers. 

\subsection{Analysis of absorption lines on quasar emission features}

In practise, most of the absorption lines from the systems considered in this paper 
are only partially spectroscopically resolved or unresolved. Furthermore, they often involve 
blends of adjacent velocity components for a given transition or blends of various transitions 
(e.g., those from C\,{\sc i} and C\,{\sc i}$^{\star}$). 
Thus, we need to rely on a line fitting procedure; we used 
VPFIT10.2\footnote{http://www.ast.cam.ac.uk/~rfc/vpfit.html} 
for this purpose, and we adopted the oscillator strength $f$  values of Morton (2003). 
If for a given system there is more than one 
transition seen against the quasar continuum, they were analysed together to obtain a fit of the 
normalised spectrum under the assumption of a uniform covering of the 
background source.  
This fit ($F_{fit}(\lambda)$) was then compared to the profile that is observed for 
the transitions that fall on emission lines. If the latter appear weaker than the 
prediction drawn from the fit, the partial covering model (with $C_f < 1$) can be used 
for these transitions. The associated corrected synthetic profile writes
\begin{equation}
F_{fit,corr}(\lambda) = C_f \, F_{fit}(\lambda) + (1 - C_f).
\end{equation}
The optimum $C_f$ value, or a lower bound when no evidence of $C_f  < 1$ is found, can be 
obtained by minimizing the $\chi^2$ computed from the difference between the observed and 
corrected synthetic profiles. 

Often, too few lines arising on the continuum alone are available; we therefore 
simultaneously fit the profile of all transitions that are observed, against both the continuum 
source and the ELR. We first assume $C_f = 1$ for all lines; if the fit is not satisfactory 
with, for instance, absorption lines seen on an emission feature that 
are overfitted, we consider 
$C_f < 1$ values for these transitions. In this case, the simultaneous fit must be performed 
after rescaling the observed spectrum near the corresponding lines 
to account for the fact that only a fraction $C_f$ of the flux is affected by absorption: 
\begin{equation}
F_{corr,n}(\lambda) = \frac{F_{obs,n}(\lambda) - (1 - C_f)}{C_f }.
\end{equation}
Again, the optimum $C_f $ value can be obtained by $\chi^2$ minimization (in this case, 
all transitions are used to compute $\chi^2$).

If the opposite applies and absorption lines detected on quasar emission features appear to be 
underfitted, the two-value model must be adopted. Once the absorption profiles 
towards the ELR have been derived from Eq. (6), they can be directly 
fitted, provided at least two 
distinct transitions occurring on emission lines are available. This leads to a separate 
determination of $N$ and $b$ parameters for the gas toward the continuum source and ELR  
(this case is illustrated in Sect. 3.2.8 by the C\,{\sc i} system at $z_{abs}=2.41837$ in 
QSO~J1439+1117).

The curve-of-growth approach can also be used if the absorption involves well-detached 
velocity components and if no blending of distinct transitions is present. The usual curve-of-growth method implicitly assumes uniform covering of the source. In the presence of partial 
covering, it follows directly from Eq. (2) that the equivalent width $W'$ becomes
\begin{equation}
W' = \int (1 - F_{obs, n}(\lambda)) \ d\lambda = 
C_f \int (1-e^{-\tau(\lambda)}) \ d\lambda = C_f \ W,
\end{equation}
where $W$ is the value that one would obtain if the source were fully covered. Alternatively, in the 
framework of the two-value model, Eq. (6) leads to 
\begin{equation}
W' = \frac{W_c+{x\,W_{elr}}}{1+x}
,\end{equation}
where $W_c$ and $W_{elr}$ are the equivalent widths that one would measure separately 
towards the 
continuum and ELR sources, respectively ($W'$ is just the flux-weighted average, as expected).
If enough transitions are available to define a curve of growth for the absorber situated in 
front of the continuum source (i.e., $W_c$ values), the location of observed $W'$ values for 
transitions that fall on quasar emission lines relative to this curve will indicate whether 
the absorption formed against the ELR is weaker or stronger than that against the continuum 
source ($W' < W_c$, thus $W_{elr} < W_c$ in the former case and $W' > W_c$, thus $W_{elr} > W_c$ 
in the latter).

\subsection{Reliable identification of non-uniformity effects}
We now discuss in which conditions it is possible to unambiguously establish the reality of 
non-uniformity effects in a quasar absorption system. An obvious difficulty comes from the
fact that partial covering of the background source and components 
with unresolved saturated features can both affect line profiles or equivalent width ratios in 
the same way.


We first consider two lines from the same species detected on a quasar emission feature and 
examine whether their relative strength can be used to assess 
non-uniform covering. If the lines are close to each other in the spectrum 
(as in the case of a Mg\,{\sc ii} doublet, for instance) and remain unresolved, 
both equivalent widths will be affected in a similar way 
(the two lines are characterised by the same $x$ and hence $C_f$ values) 
and their ratio is unchanged. 
The situation is more favorable when lines are fully resolved, especially if some display a 
flat-bottom profile, leading to a direct determination of the LFR and then of $C_f$ 
(cf. Eq. (3)). Using at 
least two lines is important to check that the flat bottom is not due to a blend of several 
unresolved velocity components, even though this is unlikely when the flat core is extended 
enough because an unrealistic combination of widths, opacities, and component separations
would be required to produce a flat profile. 
There are cases of intervening systems for which the LFR is close to, but different from, zero: 
this was first reported for H$_2$ lines (Balashev et al. 2011; see also Sect. 3.3), 
and we have discovered this effect for Fe\,{\sc ii} lines (see Sect. 3.2.4).  

Systems in which one line (line 1) is seen against the continuum source and the other (line 2) over an emission feature provide much better constraints, especially for 
systems involving (generally unresolved) lines from neutral or molecular gas, which are the 
main motivation for this paper. In the partial covering model, the equivalent widths are 
$W'_1 = W_1$  (we assume $C_c = 1$) 
and $W'_2 = C_f W_2$, where $W_1$ and $W_2$ are the equivalent width values 
expected for a fully covered source, and their ratio is 
\begin{equation}
r' = \frac{W'_2}{W'_1} = C_f \frac{W_2}{W_1}.
\end{equation}
If instead line 1 were falling on the emission line, these relations would write
$W'_1 = C_f W_1$, $W'_2 = W_2$ , and $r' = \frac{1}{C_f} \frac{W_2}{W_1}$.
To illustrate the behaviour of $r'$, we consider the C\,{\sc i}$\lambda$1560 and 
C\,{\sc i}$\lambda$1656 transitions; depending on $z_{abs}$ and $z_{em}$, these lines can
be seen with either C\,{\sc i}$\lambda$1560 or C\,{\sc i}$\lambda$1656 appearing on 
the quasar C\,{\sc iv} emission line (see Sect. 3.2.10 for such a case). 
Assuming an emission line to continuum flux ratio, $x=0.5$,  
we plot in Fig. 2 the variation of $r'$ with 
$C_f$ for Gaussian velocity components with various opacity values, 
$\tau$(C\,{\sc i}$\lambda$1560). The useful part of the diagram corresponds to 
$C_f > \frac{1}{1+x}=0.667$, the minimum value reached when 
$C_{elr} \rightarrow 0$. For full covering, 
the $r' = W'$(C\,{\sc i}$\lambda$1656)/$W'$(C\,{\sc i}$\lambda$1560) ratio must lie in 
the range $r'_{min} < r' < r'_{max}$ with $r'_{min} = 1.061$ (the optically thick limit, 
for which $W$ scales as $\lambda$) and $r'_{max} = 2.169$ (the thin limit: 
$W \propto \lambda^2 f$), which remains true even if unresolved components are present. 
Thus, if $r'$ values that fall outside this interval are measured, this is necessarily
a signature of non-uniform covering. As can be seen in Fig. 2, $r'< r'_{min}$ can be 
observed for relatively opaque lines when C\,{\sc i}$\lambda$1656 appears on the emission 
line (thick blue lines), whereas  $r'> r'_{max}$ is found for moderate opacities when 
C\,{\sc i}$\lambda$1560 coincides with the emission. 
   \begin{figure}
\hspace{0mm}
        \includegraphics[width=9.5cm,angle=0]{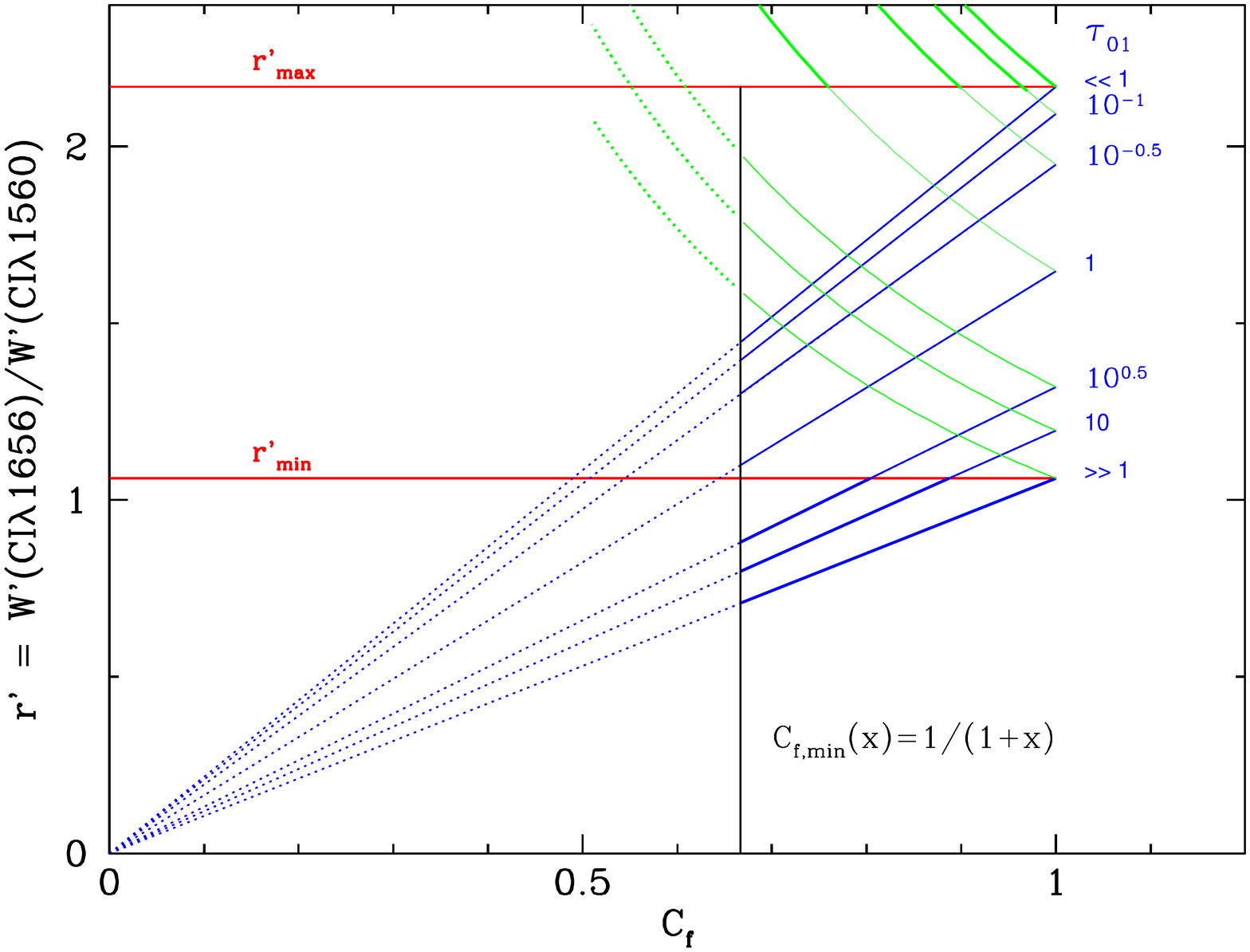}
    \caption{Variation in equivalent width ratio 
$r' = W'$(C\,{\sc i}$\lambda$1656)/$W'$(C\,{\sc i}$\lambda$1560) with $C_f$
for several values of the C\,{\sc i}$\lambda$1560 opacity (indicated in the 
upper right of the plot) and assuming $x=0.5$. Blue and green lines correspond to C\,{\sc i}$\lambda$1656 or C\,{\sc i}$\lambda$1560,
respectively, coinciding with 
the quasar emission line. The part of the plot for which $C_f < C_{f, min}$, 
assuming $C_c=1.0$,  is unphysical (dotted lines). 
Thick portions of lines correspond to values of $r'$ 
that can be exclusively obtained as a result of non-uniform covering.
}       
   \label{PB2}  
   \end{figure} 
A similar plot could be made in the two-value model, using $\tau_{elr}/\tau_c$ to 
quantify the non-uniformity of the absorber in front of the quasar source. For each 
transition $i$ ($i=1$ or 2), one can define the equivalent widths $W_{c,i}$, and 
$W_{elr,i}$; the $r'$ ratio takes the form
\begin{equation}
r'=\frac{W'_2}{W'_1} = \frac{W_{c,2}}{W_{c,1}}\frac{1+x\frac{W_{elr,2}}{W_{c,2}}}{1+x},
\end{equation} 
assuming that line 2 falls on the quasar emission line. $\frac{W_{elr,2}}{W_{c,2}}$
can be inferred from the $\tau_{elr}/\tau_c$ value, provided some assumption is made about 
the respective $b$ values for the gas lying in front of the ELR and continuum source 
(e.g., $b_{elr}=b_c$).

Still better constraints can be obtained when several transitions are available to model the 
absorption profile towards the continuum source. An ideal - but rare - 
situation occurs when two distinct transitions from the same species are seen on quasar emission 
lines, allowing us to check the consistency of the analysis, especially if $C_f < 1$ or  
 $\tau_{elr} \neq \tau_c$.  

\section{Absorber sample and analysis}

\subsection{Sample of quasars  and their absorption systems}
The UVES and HIRES data for the quasars we selected to investigate spatial covering of intervening 
absorbers  are all public and have high S/N spectra: UVES (Bergeron et al. 2004; Molaro et al. 
2013) and HIRES (Prochaska et al. 2007). \\
First, we selected quasar absorption systems that trace molecular and cold neutral gas; the species 
we considered are mainly H$_2$, C\,{\sc i,} and Fe\,{\sc i}. The systems of greatest interest are those 
where some absorption lines fall on the background quasar emission lines and for species with 
enough transitions to provide good constraints for line profile fitting. \\
We then extended our study to a few systems involving moderately ionised gas, 
mainly C\,{\sc ii}, 
Fe\,{\sc ii}, Ni\,{\sc ii,} and Si\,{\sc ii},  for which we wish to check for the absence of 
substructure at pc scales. Finally, we included the few cases for which the interstellar medium, 
local or at low redshift, could be studied by its Ca\,{\sc ii} absorption falling on Ly$\alpha$ 
or C\,{\sc iv} quasar emission lines. \\
The quasars under investigation are listed in Table 1. Hereafter, the concordance cosmological 
model is adopted.  

\begin{table} 
\caption[]{The sample.} 
\begin{center}
\begin{tabular}{l@{\hspace{3.5mm}}l@{\hspace{2.5mm}}c@{\hspace{4.5mm}}r@{\hspace{3.5mm}}c}
\hline  
\noalign{\smallskip}     
target  & $z_{\rm em}$ & spectrograph & $\Delta t$ & date  \\    
name$^a$ & & & hr  &      \\ 
\noalign{\smallskip}    
\hline  
\noalign{\smallskip}
HE 0001$-$2340 & 2.280 & UVES & 12.0 &  06-08/2001   \\
 & &  UVES &  15.0 & 09/2009  \\
\noalign{\smallskip}\hline\noalign{\smallskip}
PKS 0237$-$23 & 2.225 & UVES & 25.3 & 2001-2002 \\
 & &  UVES & 18.8 & 2011-2013  \\
\noalign{\smallskip}\hline\noalign{\smallskip}
Tol 0453$-$423 & 2.261 & UVES & 16.2 & 01/2002   \\
 & &  UVES & 16.8 & 03-11/2011  \\
\noalign{\smallskip}\hline\noalign{\smallskip}
TXS 1331$+$170 & 2.089 & UVES & 8.5 & 03-04/2011   \\
 & & HIRES($>$4220\AA) & 10.0 & 04-06/1994    \\
\noalign{\smallskip}\hline\noalign{\smallskip}  
QSO J1439+1117 & 2.583 & UVES & 8.2 &  03/2007   \\
\noalign{\smallskip}\hline\noalign{\smallskip}  
PKS 1448$-$232  & 2.208 & UVES & 13.5& 06-07/2001 \\
\noalign{\smallskip}\hline\noalign{\smallskip} 
FBQS J2340$-$0053 & 2.085 & UVES  & 7.5 & 10/2008    \\
& &  HIRES & 4.2  & 08/2006  \\
\noalign{\smallskip}    
\hline  
\noalign{\smallskip}
\multicolumn{5}{l}{$^a$ : Resolved by SIMBAD.} \\
\end{tabular}   
\end{center}
\label{list}    
\end{table}
%

        
   \begin{figure}
   \centering   
       \includegraphics[width=6.5cm,angle=0]{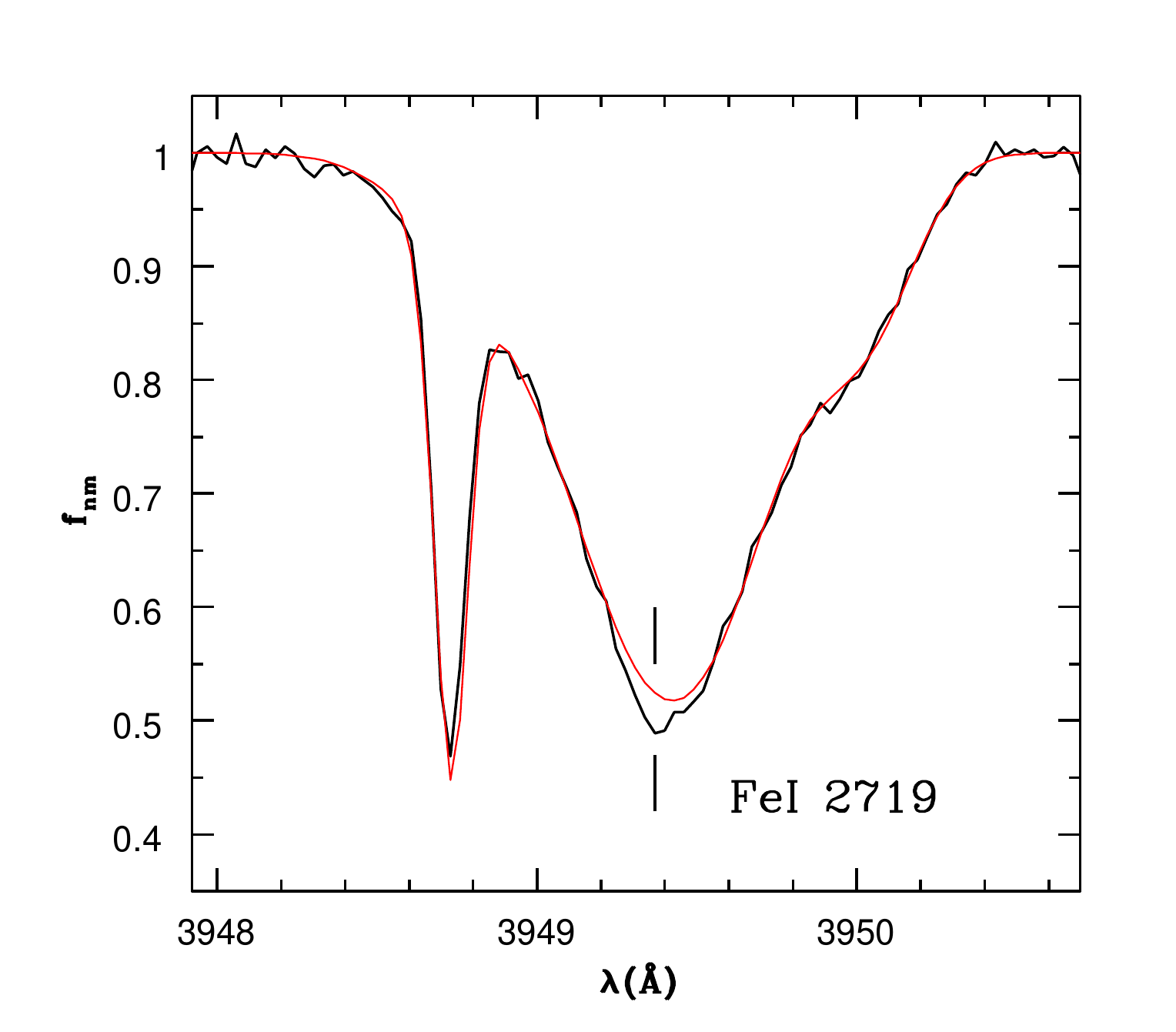}
    \caption{Intervening Fe\,{\sc i}$\lambda$2719 absorption at $z_{abs} = 0.45206$ (as marked) 
     towards 
     the quasar HE0001$-$2340. This line falls on the blue wing of the quasar Ly$\alpha$ 
     emission line. The spectrum (black curve) is shown together with a fit (see Sect. 3.2.3) 
    of three blended Ly$\alpha$ absorptions, and bluewards,  one Si\,{\sc ii} 
    absorption  at $z_{abs}=1.58643$ (red curve). 
    The Fe\,{\sc i}$\lambda$2719 absorption is unexpectedly very weak (see  Fig. 4).  
}       
   \label{He0001FeILya} 
   \end{figure} 
   \begin{figure}
   \centering   
        \includegraphics[width=10.5cm,angle=0]{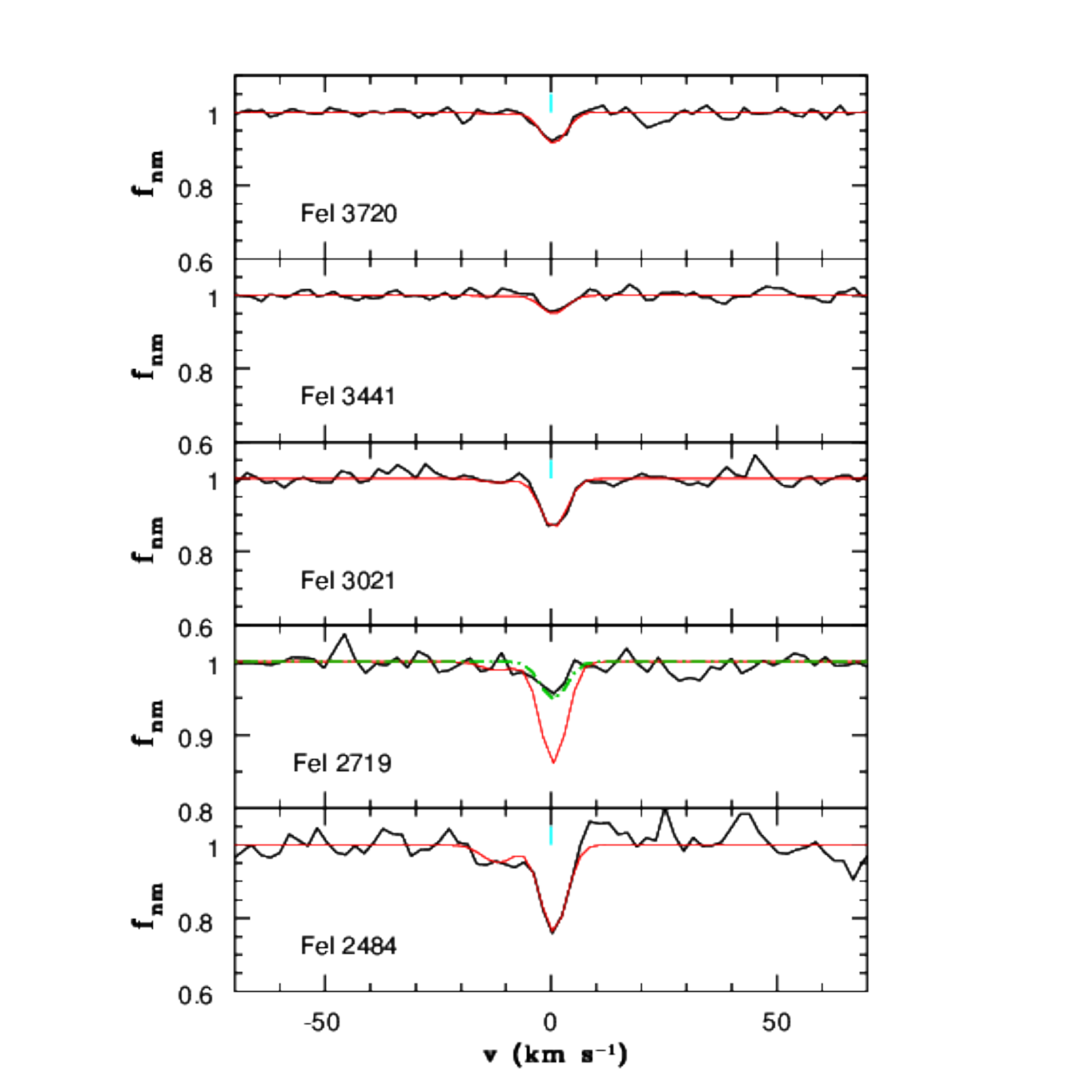}
    \caption{Intervening Fe\,{\sc i} absorption towards the quasar HE0001$-$2340: 
    UVES 2001 spectrum (black curve) and simultaneous fit for the four transitions  
    falling on the quasar continuum (red curve), thus 
    excluding the Fe\,{\sc i}$\lambda$2719 line, which falls on the quasar 
    Ly$\alpha$ emission line.  The green curve for the Fe\,{\sc i}$\lambda$2719
    absorption corresponds to a fit with a spatial coverage factor $C_{f,max}=0.37$.  
   At $v_{\rm helio}=0$~km~s$^{-1}$, the redshift is $z = 0.452060$. 
}       
   \label{He0001FeI}    
   \end{figure} 
%
   \begin{figure}
        \includegraphics[width=7.cm,angle=-90]{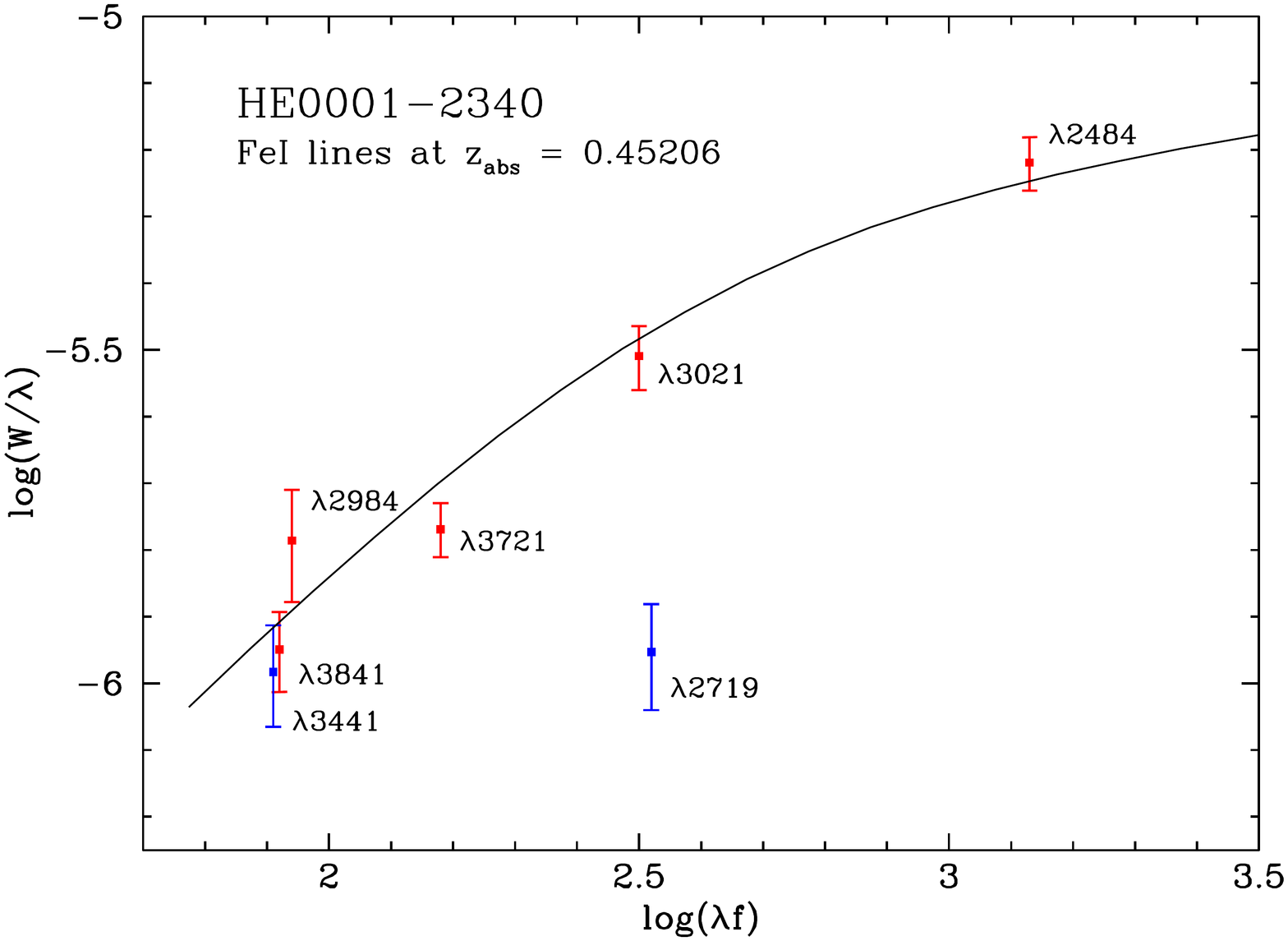}
    \caption{Curve of growth for the Fe\,{\sc i} transitions detected in the $z_{abs} = 0.45206$
 system towards HE 0001$-$2340. For each point we indicate the rest wavelength 
of the Fe\,{\sc i} transition involved. The smooth curve corresponds 
to $b = 0.55$ km~s$^{-1}$ and log $N$(Fe\,{\sc i}) = 12.289 (see text). The 
Fe\,{\sc i}$\lambda$2719 line falls near the top of the quasar Ly$\alpha$ 
emission and is characterised by a high value of the ELR to continuum flux ratio
 ($x = 2.31$). We note that the corresponding $W$ measurement lies well below 
 the curve-of-growth expectation because of partial covering of the ELR. 
The Fe\,{\sc i}$\lambda$3441 is also characterised by a low $x$  
value, and a barely significant shift is seen. 
}       
   \label{He0001cog}    
   \end{figure} 

\subsection{Metal systems with absorption line(s) on the quasar Ly$\alpha$ 
or C\,{\sc iv} emission lines} 

In this subsection, we successively describe metal systems involving 
either neutral gas (as traced by C\,{\sc i}, Fe\,{\sc i,} and Ca\,{\sc ii}) or 
moderately ionised gas. Systems involving diffuse molecular gas are 
considered in Sect. 3.3. 

\subsubsection{ HE 0001$-$2340: the Ca\,{\sc ii} system at $z_{\rm abs}=0.2705$ }

The Ca\,{\sc ii}$\lambda$3934,3969 doublet falls on the blue wing of the C\,{\sc iv} 
emission line. A differential covering effect is thus possible since the ratios of 
the emission line to the quasar continuum flux differ for the two transitions: 
$x$(Ca\,{\sc ii}$\lambda$3934)$=0.36$ and $x$(Ca\,{\sc ii}$\lambda$3969)$=0.86$. 
This is a simple system, with a main component of moderate strength  at 
$z_{\rm abs}=0.270515$, partly resolved, and a very weak component blueshifted by $-19.4$ km 
s$^{-1}$. The fit is good and there is no suggestion of any spatial covering effect; for 
the main component we obtain $N$(Ca\,{\sc ii}) $= (3.71\pm0.12)\times 10^{11}$ cm$^{-2}$ and 
$b=3.38\pm0.35$ km s$^{-1}$. 
 \\
In the red part of the spectrum, the associated Na\,{\sc i}$\lambda$5891,5897 doublet is 
not detected. In the  Ly$\alpha$ forest, there is an associated strong multiple component 
Mg\,{\sc ii} absorption doublet as well as a Fe\,{\sc ii}$\lambda$2586,2600 doublet 
(partly blended). 

\subsubsection{ HE 0001$-$2340: the Fe\,{\sc i} system at $z_{\rm abs}=0.45206$ }

This very peculiar absorber has been studied by D'Odorico (2007) and Jones et al. (2010), 
with extensive photoionisation modelling. There are 
associated absorptions by rare neutral 
species, Si\,{\sc i} and Ca\,{\sc i}, and these authors concluded that this neutral gas system 
traces a cold medium ($T\lesssim 100$ K) of high density ($n_H \sim 30-1000$ cm$^{-3}$). 
There are two available UVES spectra, taken about eight years apart (see Table 1); they both have 
very good S/N, with a higher S/N redwards of Ly$\alpha$ emission for the 2001 
spectrum. Our analysis is based on the latter, which was also the spectrum 
used by D'Odorico and Jones et al. in their studies of this system. 
\\
To estimate the column density $N$ and line width $b$ of the $z_{\rm abs}=0.45206$ Fe\,{\sc i} 
absorber, we have selected the three stronger well-detected transitions
that fall on the quasar continuum in the UVES 2001 spectrum: Fe\,{\sc i}$\lambda$3021,3720
 redwards of Ly$\alpha$ emission, and Fe\,{\sc i}$\lambda$2484 
(unblended line) in the Ly$\alpha$ forest. 
A good fit is achieved with a single, unresolved component together with  full coverage 
$C_f=C_c=1$, and we get $N$(Fe\,{\sc i}) $= (1.95\pm0.25)\times 10^{12}$ cm$^{-2}$ and 
$b=0.55\pm0.05$ km s$^{-1}$. 
\\
There is a  very weak Fe\,{\sc i}$\lambda$2719 absorption that falls on the blue side of 
Ly$\alpha$ emission, in a region  where a blend of three  Ly$\alpha$ absorptions 
($z_{\rm abs}=2.24825$, 2.24876 and 2.24924) is present together with,  
 bluewards, one Si\,{\sc ii}$\lambda$1526 ($z_{\rm abs}=1.58643$) absorption. 
A plot of the normalised spectrum of this region is shown in Fig. 3, 
with a fit of the three Ly$\alpha$ and the Si\,{\sc ii} absorptions. 
The Fe\,{\sc i}$\lambda$2719 absorption is unexpectedly very weak, although its oscillator 
strength $f$ is similar to that of Fe\,{\sc i}$\lambda$3021  (see  Fig. 4). This points 
towards a strong spatial covering effect for this Fe\,{\sc i} absorber. 
\\
   \begin{figure}
\hspace{-2mm}
        \includegraphics[width=7.cm,angle=-90]{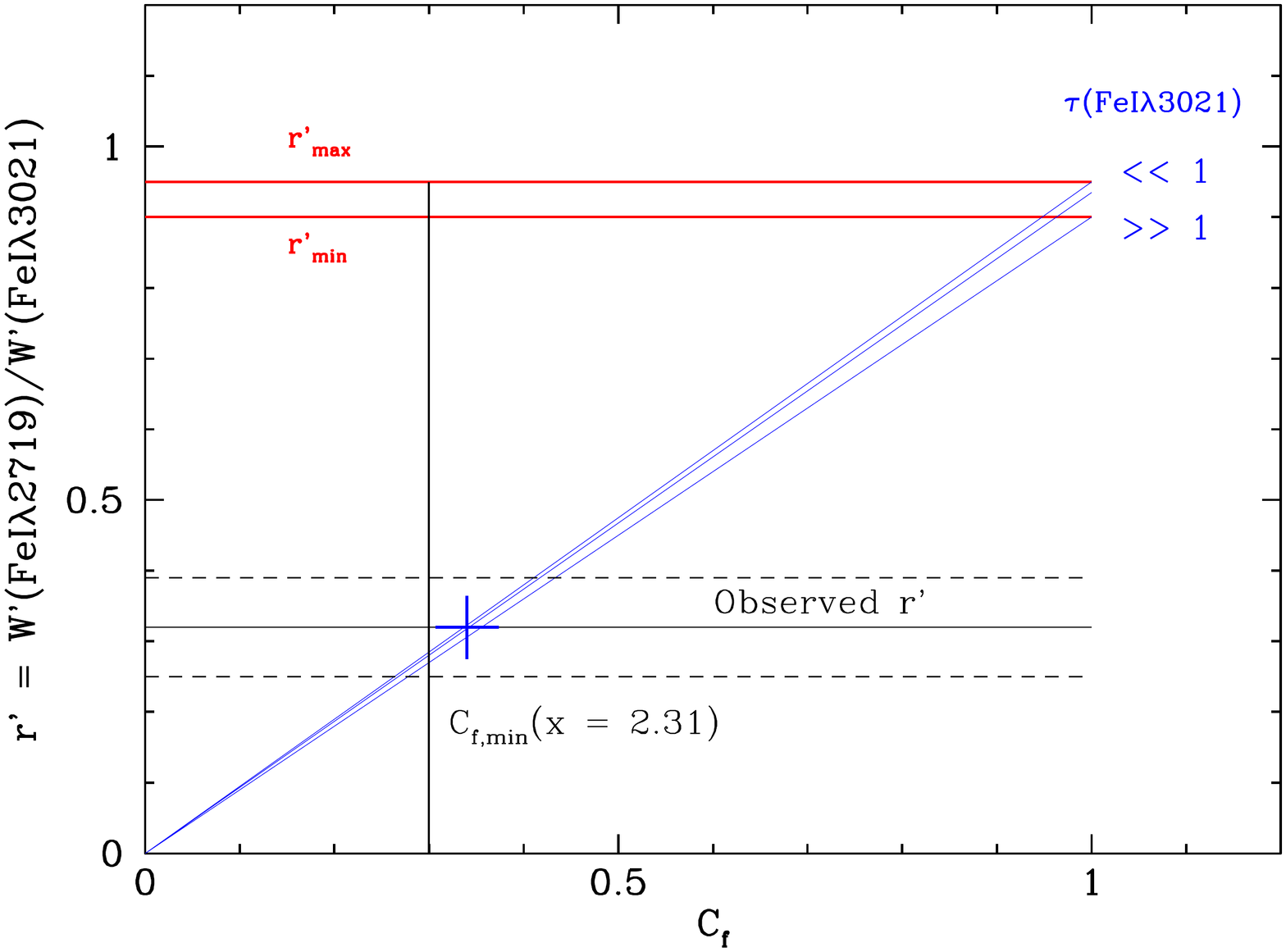}
    \caption{Same as in Fig. 2, but for the two Fe\,{\sc i}$\lambda$2719 and 
Fe\,{\sc i}$\lambda$3021 lines seen at $z_{abs} = 0.4521$ towards HE 0001$-$2340. 
For $C_f=1$, the range allowed for $r'$ is very small (red horizontal lines). 
The behaviour of $r'$ is shown for $\tau _0$(Fe\,{\sc i}$\lambda$3021) = 1, $<<1$ 
and $>>1$ (thin blue lines); $r'$ depends very little on line opacities. The  observed 
 ratio, $r' = 0.32 \pm 0.07$ (plain black horizontal line, with $1\sigma$ 
uncertainties shown as dotted lines) is clearly outside the allowed range. It implies 
a value $C_f \approx 0.33$ (thick blue cross), compatible with $C_f = C_{f,min}$
thus $C_{elr} = 0$ (thick black vertical line), irrespective of 
$\tau _0$(Fe\,{\sc i}$\lambda$3021).}   
   \label{He0001W}      
   \end{figure} 
To estimate the spatial coverage factor $C_f$, we renormalised the spectrum 
around the Fe\,{\sc i}$\lambda$2719 line, taking the Ly$\alpha$ and 
Si\,{\sc ii} absorptions mentioned above into account to derive the local continuum in this region. 
This procedure is legitimate because Ly$\alpha$ forest clouds are known to
be very large and to display no internal structure at scales comparable to
the ELR extent (Rauch et al. 2001). 
 We then used the values of $N$ and $b$ as determined with the three transitions 
falling on the quasar continuum  to fit  the renormalised spectrum around the 
Fe\,{\sc i}$\lambda$2719 line for different values of $C_f$;  the 
best values of $C_f$ correspond to the minimum of $\chi^2$. 
It should be noted that the $\chi^2$ of the fit is sensitive to values of the  
continuum rms, the latter being  inversely proportional to $C_f$ (cf. Eq. (9)).  
To minimize the effect of the continuum noise on the $\chi^2$ of the fit
around the Fe\,{\sc i}$\lambda$2719 absorption, we  therefore
limited the selected wavelength 
range around this line since the $C_f$ values are low. 
We then obtained $C_f=0.32$ and estimate that the uncertainty on this value is about 
0.05. 
\\
The determination of the coverage factor of the emission line region $C_{elr}$ requires an 
estimate of the quasar continuum flux $F_c$ underlying the Ly$\alpha$ emission line. There 
is no low-resolution spectrum of this quasar available at any epoch. Since 
variability of Ly$\alpha$ emission flux is detected even in high-redshift quasars (see, e.g., 
Woo et al. 2013: SDSS data), we used the 2001 UVES spectrum itself, which was taken in good 
seeing and clear sky conditions, to measure $F_c$ and accordingly the flux ratio $x$. We obtained $x=2.31,$ 
which implies $C_{f, min} = 0.30$ when assuming full coverage of the quasar continuum  
($C_c=1$ in Eq. 4). 
 Thus the coverage factor of the ELR is very small: it is consistent with zero,   
while the coverage factor of the quasar continuum is fully compatible with $C_c=1$.   
The maximun possible value of $C_{elr}$ is determined by $C_{f, max} = 0.37$ 
and equals $C_{elr, max}=0.10$. 
\\
The  Fe\,{\sc i} curve of growth is shown in Fig. 5, adopting the $b$ and log N values derived 
from the three transitions seen on the continuum alone. 
Some weak lines have been included 
in addition to those used to obtain the fit displayed in Fig. 4 (Fe\,{\sc i}$\lambda$2984 and 
Fe\,{\sc i}$\lambda$3841 together with Fe\,{\sc i}$\lambda$3441) in order to better sample the 
low-opacity end of the curve. The Fe\,{\sc i}$\lambda$2719 line clearly falls below the curve 
of growth: the inferred $C_f$ value, $C_f=0.33$, is fully consistent with the value obtained 
from the fit of the  Fe\,{\sc i}$\lambda$2719 absorption profile. The Fe\,{\sc i}$\lambda$3441 
line arises on the blue wing of the C\,{\sc iv} emission line. For this feature, the ELR to 
continuum flux ratio is low, $x=0.22$, which implies a $C_f$ value of  0.82 (assuming 
$C_{elr} \simeq 0$ and $C_c =1$ as for the Ly$\alpha$ ELR). This
is close enough to 1 
to account for the absence of a significant departure from the curve of growth.
\\
In Fig. 6 we show the equivalent width ratio $r'$ versus $C_f$ in the partial covering models 
for the two Fe\,{\sc i}$\lambda$2719 and Fe\,{\sc i}$\lambda$3021 transitions. The latter 
have very similar $\lambda f$ values, implying that i) the allowed $(r'_{min} ,  r'_{max})$ 
range for $C_f=1$ is very small, and that ii) $r'$ depends very little on line opacities (all the 
curves are nearly coincident, regardless of the value of  $\tau _0$(Fe\,{\sc i}$\lambda$3021)). 
The observed ratio, $r' = 0.32 \pm 0.05$, 
lies well below the possible range for $C_f = 1$, showing unambiguously that $C_f < 1$ for this 
system; the inferred $C_f$ value (corresponding to the blue cross in Fig. 6, where the line 
$r'=0.32$ intersects theoretical curves) is close to the minimum covering factor associated  
with $C_{elr} = 0$, in agreement with the optimal value derived from line fitting. This figure 
illustrates that detecting two transitions with similar opacities, one over a quasar emission
line and the other against the continuum alone, is a powerful way to establish the reality 
of partial covering effects.\\
The Fe\,{\sc i}$\lambda$3441 transition, which is on the C\,{\sc iv} emission line, is 
intrinsically too weak to detect a significant partial covering effect. As mentioned above, its 
W measurement lies barely below the curve of growth expectation for full spatial covering.  In addition,  
nearly equally good fits were obtained when we added this transition to the three that fall on the 
quasar continuum assuming either $C_f = 1$ or 0.88 (i.e., $C_{elr} = 0.10$); these fits have very 
similar values of  $N$(Fe\,{\sc i}) and $b$ (the difference is about 5 \%).  

The Mg\,{\sc ii} doublet associated with the Fe\,{\sc i} system at 
$z_{\rm abs}=0.45206$ falls on the red wing of the N\,{\sc v} emission line. Jones et al. 
(2010) found evidence of partial covering in this Mg\,{\sc ii} system and estimated 
$C_f  \approx 0.6$. As discussed in Sect. 2, spatial covering effects are difficult to 
ascertain in such an unsaturated system because both transitions fall
in a small wavelength range and thus have nearly the same associated $x$ values
(we estimate $x=0.92$ at Mg\,{\sc ii}$\lambda$2796).
Furthermore, since the LFR and $W$ ratio lie in the allowed range for $C_f = 1$, it should
be possible to obtain an acceptable fit that is consistent with $C_f = 1$. Using the 
procedure described above, we obtain a good fit of this doublet assuming full coverage :
$N$(Mg\,{\sc ii}) $= (1.41\pm0.03)\times 10^{12}$ cm$^{-2}$ and $b=2.50\pm0.10$ km s$^{-1}$
(which corresponds to $\tau _0$(Mg\,{\sc ii}$\lambda$2796) = 1.5). 
Equally good fits are achieved down to $C_f=0.75$, which 
leads to a possible range of the ELR coverage factor, $0.48 \lesssim C_{elr} \leq 1.0$.


\begin{table*} 
\caption[]{Systems with absorption line(s) on the quasar Ly$\alpha$ or C\,{\sc iv} emission.}    
\begin{center}
\begin{tabular}{l@{\hspace{3.5mm}}l@{\hspace{2.5mm}}c@{\hspace{4.5mm}}c@{\hspace{3.5mm}}c@{\hspace{3.5mm}}c@{\hspace{3.5mm}}c@{\hspace{5.5mm}}c@{\hspace{3.0mm}}c}
\hline  
\noalign{\smallskip}     
target  & $z_{\rm em}$ &  $z_{\rm abs}$ & element & number & number  &  structure & $C_f$ & $C_{elr}$ \\    
name & & &   & transitions  &  components &  effects &  &   \\ 
\noalign{\smallskip}    
\hline  
\noalign{\smallskip}
HE 0001$-$2340 & 2.280 & 0.27052 & Ca\,{\sc ii} & 2  & 2  & no  & 1.0  & 1.0    \\
 & & 0.45206 & Fe\,{\sc i} & 5  & 1 $+ 1^{b,c}$    & yes & $0.30-0.37$  &  $0.0-0.10$   \\
 & & 0.45206 &  Mg\,{\sc ii} & 2 & 1     & possible & 0.75 - 1.0   & 0.48 - 1.0  \\
\noalign{\smallskip}\hline\noalign{\smallskip}
PKS 0237$-$23 & 2.225 & 1.36469 & C\,{\sc i} & 2  & 5 & possible  & $0.80-0.90$ & $ 0.71-0.85$ \\ 
 & &  1.36469 & Fe\,{\sc ii} & 5  & 13  & no &  1.0  & 1.0  \\
\noalign{\smallskip}\hline\noalign{\smallskip}
Tol 0453$-$423 & 2.261 & 0.72604 & Fe\,{\sc ii} & 2  & 5+4  & yes  &   0.98 &   0.96    \\
&  & 0.72604 & Mn\,{\sc ii} & 3  & 5  & no   &  1.0 &  1.0    \\
\noalign{\smallskip}\hline\noalign{\smallskip}
TXS 1331$+$170 & 2.089 & 0.74461 & Fe\,{\sc i} & 5  & 1  &   possible & $ 0.80-1.0 $  & $ 0.56-1.0$ \\ 
 & & 1.32828 & Fe\,{\sc ii} & 6  & 9  &  no   &  1.0  &  1.0  \\
 & & 1.77653 & C\,{\sc i} &  3  & 2 $+1^c$    &   possible & $0.90-1.0$  &  $ 0.66-1.0$  \\ 
\noalign{\smallskip}\hline\noalign{\smallskip}  
QSO J1439+1117 & 2.583 & 2.41837 & C\,{\sc i} & 5  &  7  & yes  & see  & text \\
\noalign{\smallskip}\hline\noalign{\smallskip} 
PKS 1448$-$232  & 2.208 & $-$0.00002 & Ca\,{\sc ii} & 2 & 10 & no   & 1.0  & 1.0 \\
\noalign{\smallskip}\hline\noalign{\smallskip} 
FBQS J2340$-$0053 & 2.085 & 2.05454 &  C\,{\sc i} & 5  & 8 & yes  &  $0.85-0.90$  & $ 0.63-0.75$     \\
\noalign{\smallskip}    
\hline  
\noalign{\smallskip}
\multicolumn{9}{l}{$^b$ : Blended components.} \\
\multicolumn{9}{l}{$^c$ : Additional component either very weak or noisy.} \\
\end{tabular}   
\end{center}
\label{obs}     
\end{table*}
%


   \begin{figure}
\hspace{-7mm}
        \includegraphics[width=10.cm,angle=0]{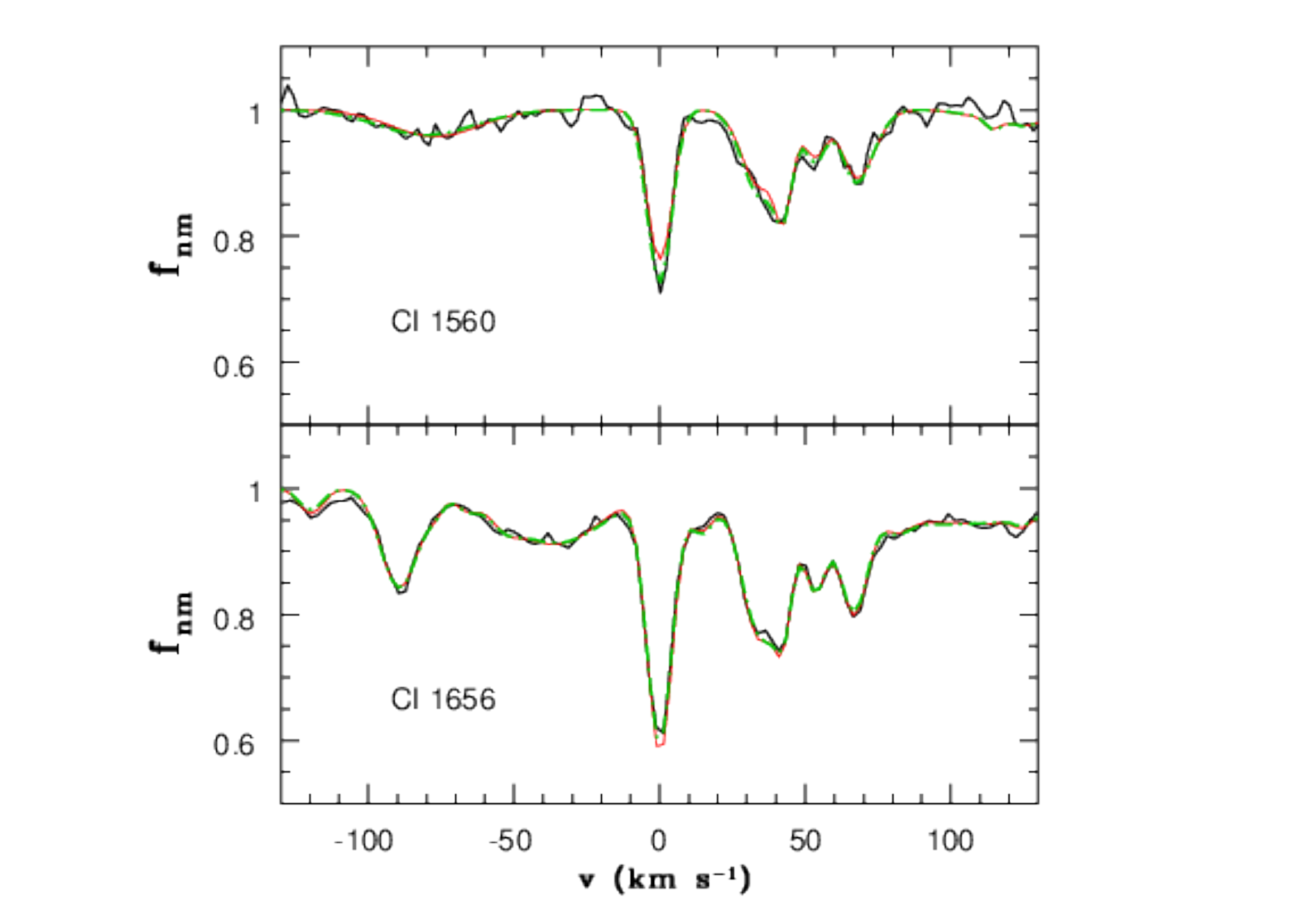} 
    \caption{Intervening C\,{\sc i} absorption towards  the quasar PKS 0237$-$23 : 
    spectrum (black curve) and simultaneous fit to the two transitions (red curve). 
    The green curves correspond to a fit with 
    a spatial coverage factor $C_f=0.85$ of the Ly$\alpha$ emission line.  
  In the bottom panel, bluewards of C\,{\sc i}$\lambda$1560 line (which falls 
  at the top of Ly$\alpha$ emission)  C\,{\sc iv}$\lambda$1550 absorption at $z_{\rm abs}=1.52580$
  is present as well as a weak, somewhat broad Ly$\alpha$ absorption. 
   At $v_{\rm helio}=0$~km~s$^{-1}$, the redshift is $z = 1.364695$. 
}
   \label{P0237CI}      
   \end{figure} 


\subsubsection{ PKS 0237$-$23: the  C\,{\sc i} system at $z_{\rm abs}=1.3650$}

Srianand et al. (2007) analysed UV (IUE) and 21 cm (GMRT) data to derive $N$(H\,{\sc i}) for this 
absorber and concluded that this system is a sub-DLA. They also derived abundances by simultaneously 
analysing (same $z_{\rm abs}$ and $b$) neutral and singly-ionised 
species; very many components were included in their fit, and the 
$b$ values they derived for the C\,{\sc i} components are in the range 2.5-6.8 km s$^{-1}$. 
\\
This is a case for which there are only two available C\,{\sc i} transitions, each with multiple 
absorption components: C\,{\sc i}$\lambda$1656, which falls at the top of Ly$\alpha$ 
emission, and C\,{\sc i}$\lambda$1560,  which is in a fairly clean region with high 
S/N of the Ly$\alpha$ forest. Our fit of the regions around these two transitions 
includes a C\,{\sc iv}$\lambda$1550 absorption at $z_{\rm abs}=1.52580$ 
bluewards of C\,{\sc i}$\lambda$1656, and a weak somewhat broad Ly$\alpha$ 
absorption at $z_{\rm abs}=2.22263$ 
(this Ly$\alpha$ absorption accounts for some of the C\,{\sc i} components included in the 
Srianand et al. paper mentioned above) 
as well as another weak somewhat broad Ly$\alpha$ absorption at $z_{\rm abs}=2.03427$, 
bluewards of C\,{\sc i}$\lambda$1560. For the UVES 2001-2002 data, the fit for 
the main isolated component at $z = 1.364695$ and full spatial covering yields 
 $N$(C\,{\sc i}) $= (9.0\pm0.2)\times 10^{12}$ cm$^{-2}$ and $b=4.27\pm0.17$ km s$^{-1}$, 
which means that the line is partly resolved. This fit is shown in Fig. 7 (red curve). 
It is not entirely satisfactory at least for the main isolated component: the transition on the 
Ly$\alpha$ emission line is overfitted and the transition on the quasar continuum 
is underfitted, which suggests partial covering. 
For the other C\,{\sc i},C\,{\sc i}$^{\star}$ components (weak and mostly blended),  
the discrepancy between the observation and the fit is not as conspicuous. 
\\
We then examined the possibility of a partial covering effect and applied a $C_f$ correction 
factor to the normalised flux of the Ly$\alpha$ emission region. 
The best fit, obtained for the minimum $\chi^2$ value,  
yields $C_f=0.85(\pm0.05)$. For the isolated component at $z_{\rm abs}=1.364695$, 
we obtain $N$(C\,{\sc i}) $= (1.05\pm0.02)\times 10^{13}$ cm$^{-2}$ and $b=4.09\pm0.15$ km s$^{-1}$. 
The difference in the $N$ values between $C_f=1$ and 0.85 is significant at the 
5$\sigma$ level.  This fit with partial covering is also good for all the other weaker 
C\,{\sc i},C\,{\sc i}$^{\star}$ components. The $C_f$ value is consistent with the value obtained 
for the more recent UVES data (2011-2013), although this spectrum has a lower S/N ratio.
The ratio of ELR to quasar continuum flux at C\,{\sc i}$\lambda$1656 equals 
$x=2.16,$ which implies a coverage factor of the ELR of about $C_{elr}\sim0.78 $.
 \\
An alternative model, which assumes full spatial covering of the ELR,  
involves a narrow additional component that is fully blended with the strong isolated component 
at $z_{\rm abs}=1.364695$. 
The fit of this blend is poorly constrained, especially since the additional component has to 
be very narrow. A possible fit is  $N$(C\,{\sc i}) $=(1.71$ and $0.54)\times 10^{13}$ cm$^{-2}$ 
and $b=(0.5$ and $6.4$) km s$^{-1}$. 
Although such a narrow component is not unrealistic, as indeed outlined below 
for the C\,{\sc i} absorber towards TXS 1331$+$170, this alternative model 
is not favoured considering the fairly high temperature ($T\gtrsim 1000$ K) 
inferred from the 
detailed analysis of Srianand et al. (2007: the discussion of the velocity range B). 


The associated  Fe\,{\sc ii} absorption is highly multiple (13 components). 
The Fe\,{\sc ii}$\lambda$1608 line is in the Ly$\alpha$ forest, the Fe\,{\sc ii} triplet is on 
the continuum redwards of the C\,{\sc iv} emission, and the 
Fe\,{\sc ii}$\lambda$2586,2600 doublet falls on the blue wing of the C\,{\sc iii}] emission line 
(the Fe\,{\sc ii}$\lambda$2600 absorption is at the top of the emission line).  
The component associated with the C\,{\sc i} absorber is of moderate strength. For the component 
at $z_{\rm abs}=1.364994$, the Fe\,{\sc ii}$\lambda$2382,2600 lines are just about saturated. 
 The fit obtained for the five transitions redwards of the Ly$\alpha$ emission 
with full spatial covering is very good for all the components. 

   \begin{figure}
\hspace{-7mm}
        \includegraphics[width=10.cm,angle=0]{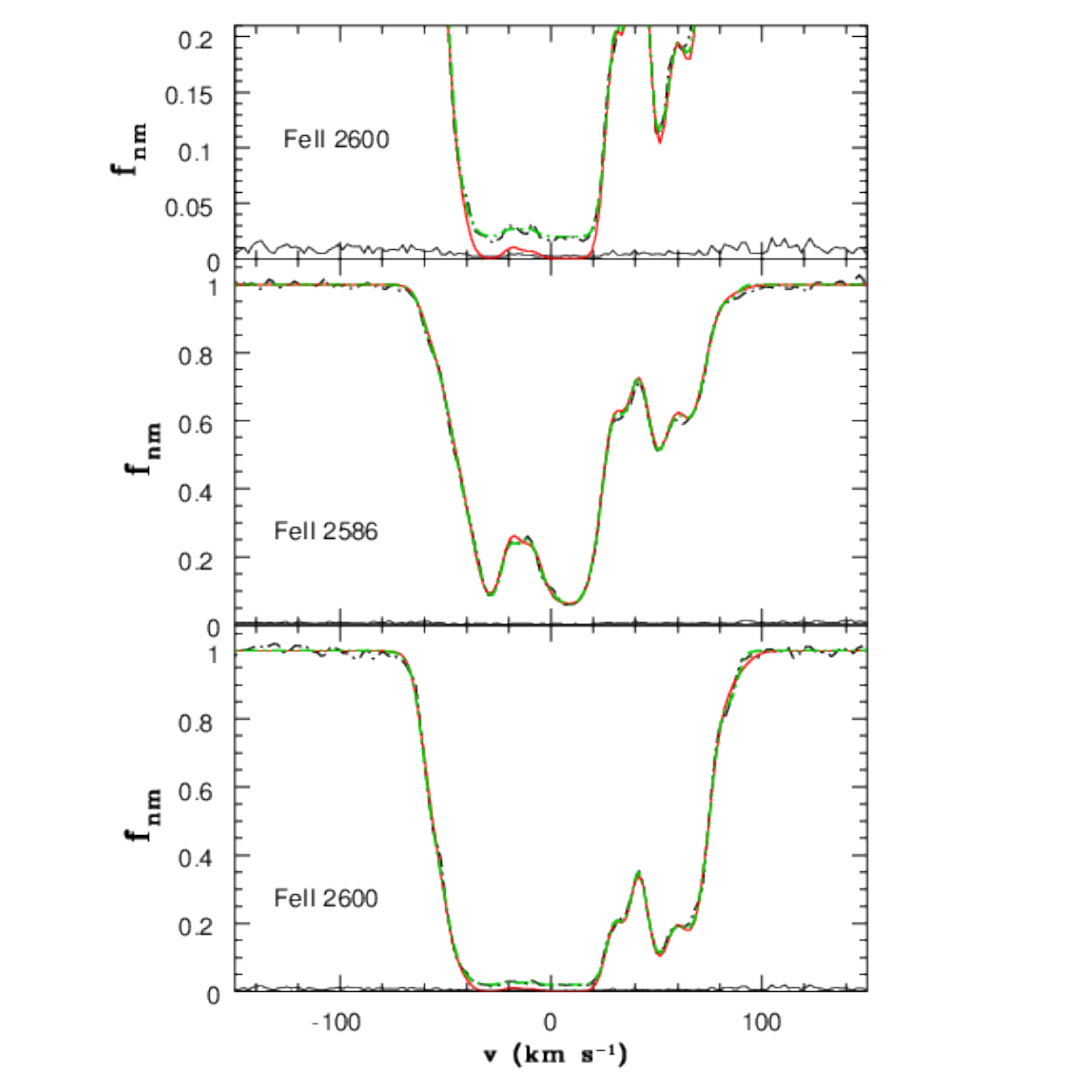}
    \caption{Intervening Fe\,{\sc ii} absorption towards the quasar Tol 0453$-$423: 
    spectrum (black curve) and simultaneous fit to the two transitions (red curve), 
    both falling on the quasar Ly$\alpha$  emission line.  
   The bottom black curve in each panel corresponds to the noise level. 
   The upper panel is a zoom of the lower to clearly show the good fit of the  
   observed spectrum as obtained with a spatial coverage factor $C_f=0.98$ (green curve). 
   At $v_{\rm helio}=0$~km~s$^{-1}$, the redshift is $z = 0.72604$. 
}       
   \label{Q0453FeII}    
   \end{figure} 

\subsubsection{Tol 0453$-$423: the Fe\,{\sc ii} system at $z_{\rm abs}=0.7261$}

This bright quasar has been extensively used to study its numerous absorption systems 
at $z_{\rm abs}>1$ (e.g., Sargent et al. 1979; Kim et al. 2013), but a detailed study of the 
$z_{\rm abs}=0.7261$ system has not yet been performed. \\
For our analysis we used the 2002 UVES spectrum, which has high S/N. 
The Fe\,{\sc ii}$\lambda$2586,2600 doublet of the $z_{\rm abs}=0.7261$ system falls on the quasar 
Ly$\alpha$-N\,{\sc v} emission. The profile of the Fe\,{\sc ii}$\lambda$2600 absorption is extremely 
unusual for a singly-ionised species: as clearly seen in Fig. 8 (top panel), 
it has a flat bottom, covering about 60 km s$^{-1}$, which does not reach the zero flux level. 
The flux residual is at the 2.0\% level; with an rms of 0.003, this yields a detection at a 
7$\sigma$ significance level.  
The fit obtained with nine components and full spatial covering is inconsistent with the data 
since the bottom of the Fe\,{\sc ii}$\lambda$2600 absorption should then reach the zero flux 
level (red curve in Fig. 8).
Although the 2011 UVES spectrum has a lower S/N, a similar residual is observed for the 
 Fe\,{\sc ii}$\lambda$2600 absorption (clearer after some smoothing of the data).
This effect is very rarely detected for singly-ionised species; another clear 
example involving Si\,{\sc ii} towards LBQS 1232$+$082 is discussed by 
Balashev et al. (2011). 
\\
The derived value of the spatial coverage factor for this Fe\,{\sc ii} absorber is $C_f=0.98$, 
which for a flux ratio $x=1.22$ gives a coverage factor of the ELR of $C_{elr}=0.96$. The 
absorber fully covers the quasar continuum since the Fe\,{\sc ii}$\lambda$2382 absorption line, 
which is in a clean part of the Ly$\alpha$ forest, reaches the zero flux level over 
100 km s$^{-1}$. 

The Mn\,{\sc ii} absorption lines associated with this Fe\,{\sc ii} system are weak with 
blended components. The Mn\,{\sc ii}$\lambda$2576 absorption line is at the top of the Ly$\alpha$ 
emission line and  Mn\,{\sc ii}$\lambda$2606 is at the knee of the Ly$\alpha$-N\,{\sc v} 
emission. Only the five components that trace the saturated part of the Fe\,{\sc ii}$\lambda$2600 
absorption line have Mn\,{\sc ii} counterparts. A good fit of the three Mn\,{\sc ii} transitions 
is obtained with $C_f=1.0$. 
Assuming that  Mn\,{\sc ii} transitions trace the same region as those of  Fe\,{\sc ii}, 
that is, adopting the same value of $C_{elr}$, we can estimate the values of $C_f$ for 
Mn\,{\sc ii}$\lambda$2576 and 2606. Despite the large difference in the values of the flux 
ratios, $x(\lambda2576,2606)=3.92,1.03$, the derived values of  $C_f$ are nearly identical 
$C_f(\lambda2576,2606)=0.97,0.98$. 
This was expected since $C_f$(Fe\,{\sc ii}) is very close to unity.

        
   \begin{figure}
\hspace{-7mm}
        \includegraphics[width=10.cm,angle=0]{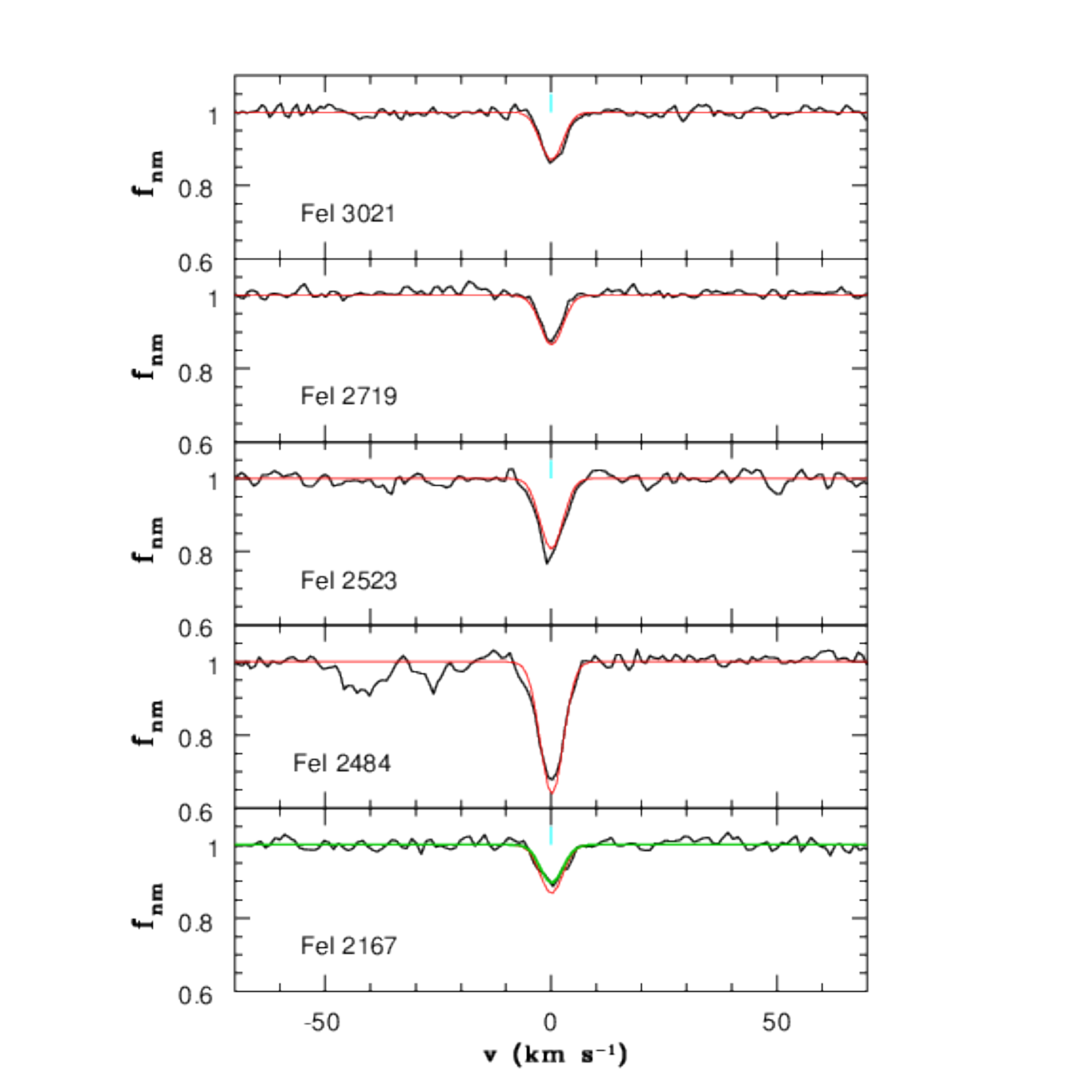}
    \caption{Intervening Fe\,{\sc i} absorption towards the quasar 
  TXS 1331$+$170:  spectrum (black curve) and simultaneous fit to the four transitions (red curve),
    excluding from the fit FeI2167, which falls on the quasar Ly$\alpha$ 
    emission line.   The green curve for the Fe\,{\sc i}$\lambda$2167 
    absorption corresponds to a fit with a spatial coverage factor $C_f=0.8$.  
   At $v_{\rm helio}=0$~km~s$^{-1}$, the redshift is $z = 0.74461$. 
}       
   \label{Q1331FeI}     
   \end{figure} 

\subsubsection{ TXS 1331$+$170: the  Fe\,{\sc i} system at $z_{\rm abs}=0.74461$}

Carswell at al. (2011) studied this quasar, but did not discuss this system. 
The 2011 UVES spectrum has high S/N  and high spectral resolution (FWHM = 5.5\AA). 
There are four well-detected Fe\,{\sc i} transitions on the quasar 
continuum as well as the Fe\,{\sc i}$\lambda$2167 absorption (S/N $\simeq 50$),
which falls on the knee of the quasar Ly$\alpha$-N\,{\sc v} emission. 
The HIRES spectrum does not cover the Ly$\alpha$ emission region, but the other 
four transitions  on the quasar continuum (Fe\,{\sc i}$\lambda$2484,2523,2719,3021) are well 
detected. The fit of these transitions involves only one component, and within the 
uncertainties,  the results are identical for the two spectra; for UVES we obtain 
$N$(Fe\,{\sc i}) $=(1.13\pm0.03)\times 10^{12}$ cm$^{-2}$ and $b=1.18\pm0.08$ km s$^{-1}$, 
and for HIRES 
$N$(Fe\,{\sc i}) $=(1.12\pm0.05)\times 10^{12}$ cm$^{-2}$ and  $b=1.06\pm0.12$ km s$^{-1}$.  
\\
The  Fe\,{\sc i}$\lambda$2167 absorption is weak (see Fig. 9) and somewhat 
overfitted with the $N$, $b$ values derived for the four transitions 
on the quasar UVES continuum. Its normalised minimum flux is also 
equal to that of the Fe\,{\sc i}$\lambda$2719 absorption, 
whereas its oscillator strength is greater than that of Fe\,{\sc i}$\lambda$2719  by 23\%. 
This suggests some spatial covering effect. From the minimum $\chi^2$ value 
of the fit we obtain $C_f\sim 0.8$; this fit is also shown in Fig. 9  (green curve). 
The flux ratio $x$ is determined from the UVES data and equals  $x=0.83$ at the position
of the Fe\,{\sc i}$\lambda$2167 absorption, implying a coverage factor of the ELR 
$C_{elr}\sim 0.56$. However, this is  uncertain since for a full covering of the 
ELR, the difference between the data and the fit is only about 1.8 times the value of 
the spectrum rms. A full covering of the ELR therefore cannot be ruled out.  


   \begin{figure}
\hspace{-7mm}
        \includegraphics[width=10.8cm,angle=0]{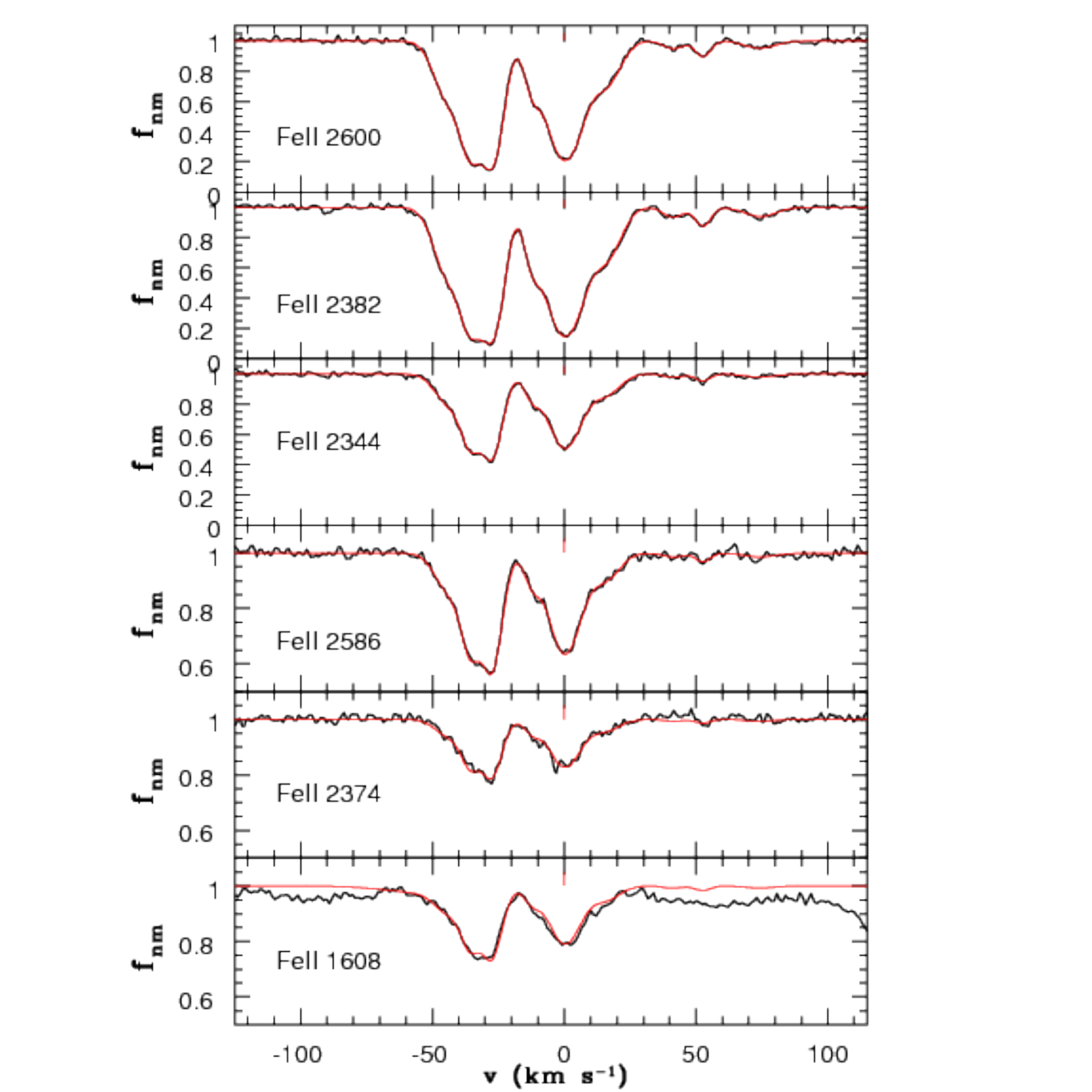}
    \caption{Intervening Fe\,{\sc ii} absorption towards the quasar TXS 1331$+$170: 
    spectrum (black curve) and simultaneous fit to the six transitions (red curve).
    The lower three panels correspond to Fe\,{\sc ii} lines of smaller 
    oscillator strength and have a different $f_{\rm nm}$ scale.  
    The Fe\,{\sc ii}$\lambda$1608 absorptions are 
    redshifted on top of the quasar Ly$\alpha$ emission, and there are weak 
    Ly$\alpha$ absorptions not included in the fit. 
   At $v_{\rm helio}=0$~km~s$^{-1}$, the redshift is $z = 1.328522$.  
}       
   \label{Q1331FeII}    
   \end{figure} 

   \begin{figure}
\hspace{2mm}
        \includegraphics[width=8.8cm,angle=0]{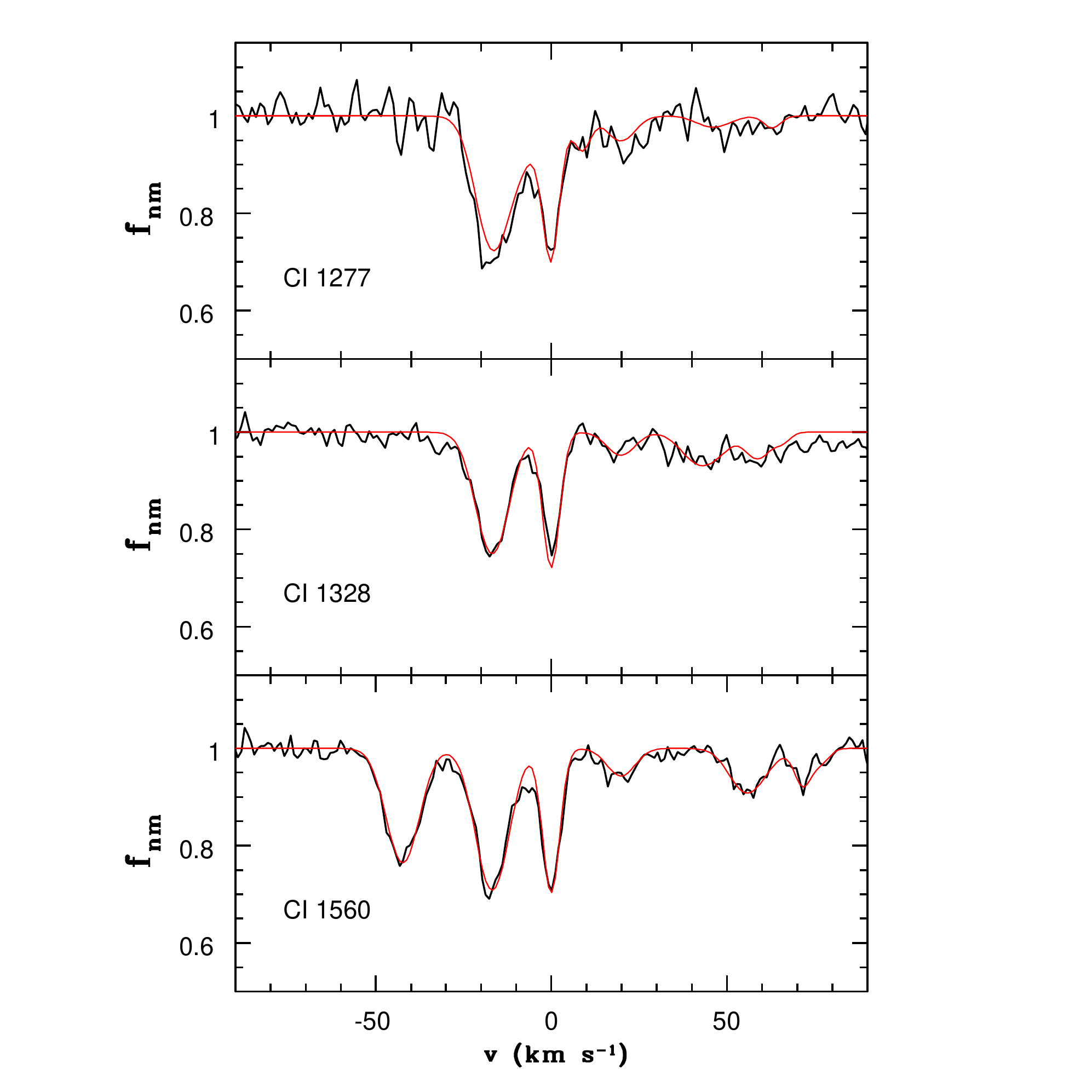}
    \caption{Intervening C\,{\sc i} absorption towards the quasar TXS 1331$+$170 : 
    spectrum (black curve) and simultaneous fit to the three transitions (red curve). 
   In the lower panel, bluewards of C\,{\sc i}$\lambda 1560,$   
   Al\,{\sc iii}$\lambda$1862 absorption is detected at $z_{\rm abs}=1.325342$.
   At $v_{\rm helio}=0$~km~s$^{-1}$, the redshift is $z = 1.776525$: this is the most narrow C\,{\sc i} 
   component ($b=0.68$ km s$^{-1}$). }  
   \label{Q1331CI}      
   \end{figure} 

\subsubsection{ TXS 1331$+$170: the Fe\,{\sc ii} system at $z_{\rm abs}=1.32828$}

The Fe\,{\sc ii}$\lambda$1608 absorption line of the $z_{\rm abs}=1.32828$ system is
redshifted on top of the quasar Ly$\alpha$ emission (only covered by the UVES spectrum). 
The other five strong  Fe\,{\sc ii} lines are on the quasar continuum and well detected 
in the UVES spectrum (there is a defect in the HIRES data at the position 
of Fe\,{\sc ii}$\lambda$2382). At the position of the Fe\,{\sc ii}$\lambda$1608 
absorption, the flux ratio equals $x\simeq1.0$. 
There are two weak Ly$\alpha$ absorption lines, one on each side of this line. 
With the fit obtained for full spatial covering and nine blended subcomponents,  
the Fe\,{\sc ii}$\lambda$1608 absorption is underfitted. 
A most likely explanation is some blending with a weak Ly$\alpha$ absorption.  
A good fit is indeed obtained with the addition of a weak blended  Ly$\alpha$ absorption component.

\subsubsection{ TXS 1331$+$170: the C\,{\sc i} system at $z_{\rm abs}=1.7764$}

The neutral and singly-ionised species of this system were analysed in detail 
by Carswell at al. (2011), including H\,{\sc i} (optical and 21 cm data), H$_2$ , and C\,{\sc i}. 
Old UVES (2002 and 2003) and HIRES data taken at different epochs were combined for this purpose. 
One of the three detected C\,{\sc i} components (at $z_{\rm abs}=1.77653$) is very narrow, 
with $b=0.55\pm0.13$ km s$^{-1}$, thus of low kinetic temperature, as confirmed by a curve-of-growth analysis. 
These authors discussed the possibility that this C\,{\sc i} cold absorber only partially 
covers the background source; they concluded that it is most unlikely since the 
corresponding saturated H$_2$ transitions have flat cores with zero residual intensities. 
We note that this is also the case for O\,{\sc i}$\lambda$1302 in the Ly$\alpha$ forest 
and C\,{\sc ii}$\lambda$1334 on the blue side of the Ly$\alpha$ emission line. 
\\
Only three transitions are well detected in the 2011 UVES spectrum:  
C\,{\sc i}$\lambda$1277 is in a clean region of the Ly$\alpha$ forest,  
C\,{\sc i}$\lambda$1328 is on the blue wing of Ly$\alpha$ emission, and 
C\,{\sc i}$\lambda$1560 is on the weak Si\,{\sc iv}  emission. 
The 1994 HIRES spectrum only covers the C\,{\sc i}$\lambda$1560,1656 transitions. 
It has a somewhat lower resolution (FHWM$=6.25$ km s$^{-1}$) than the 2011 UVES 
spectrum (FHWM$=5.5$ km s$^{-1}$); we therefore did not combine these spectra for data analysis.  
For a full spatial covering of the background source, we find a low $b$ value for 
the $z_{\rm abs}=1.77653$ component: $b=0.68\pm0.18$ and $0.82\pm0.49$ km s$^{-1}$ 
for the UVES and HIRES spectra, respectively, both with about the same column density 
$N$(C\,{\sc i}) $=(1.10\pm0.18)\times 10^{13}$ cm$^{-2}$. 
This is  consistent with the results of Carswell et al. The UVES data and their fit are shown 
in Fig. 10; the fit includes  
Al\,{\sc iii}$\lambda$1862 absorption at $z_{\rm abs}=1.325342$
bluewards of C\,{\sc i}$\lambda$1560. 
\\
Constraints on the spatial covering of the ELR by the cold component can only be obtained 
from the UVES data; we note that the spatial coverage factor should be close to unity since 
the difference between the observations and the fit  ($\sim$1.7 times the rms)
for the C\,{\sc i}$\lambda$1328 absorption only indicates a slight overfitting.  
Acceptable fits can also be obtained with a coverage factor for the Ly$\alpha$ emission region 
$C_f\neq 1$  provided that $C_f\gtrsim 0.9$ for the C\,{\sc i} component at 
$z_{\rm abs}=1.77653$.  
The flux ratio at the position of the C\,{\sc i}$\lambda$1328 transition equals $x=0.41,$ 
which implies a minimum coverage factor of the ELR $C_{elr}\gtrsim 0.66$. 
 
Four Ni\,{\sc ii} transitions associated with the C\,{\sc i} system 
are detected at $z_{\rm abs}=1.7764$. The Ni\,{\sc ii}$\lambda$1370 absorption falls on the quasar 
Ly$\alpha$-N\,{\sc v} emission line and Ni\,{\sc ii}$\lambda$1709 on the blue wing of the 
C\,{\sc iv} emission. One medium-weak component is located at $z_{\rm abs}=1.77640$ and three 
weak components redwards of this. A good fit is obtained for full spatial 
covering.

   \begin{figure}
\hspace{-7mm}
        \includegraphics[width=11.cm,angle=0]{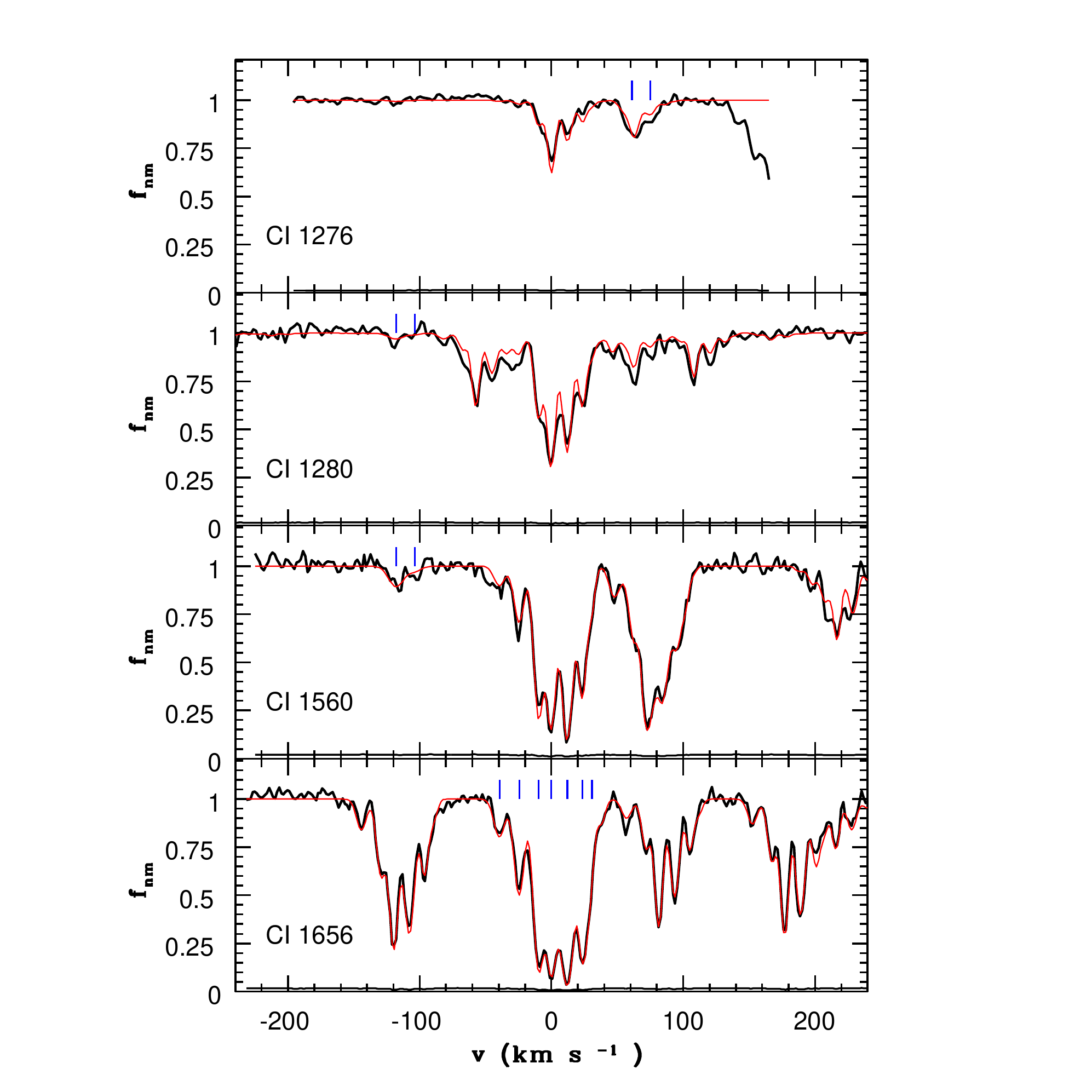}
    \caption{Intervening C\,{\sc i} absorption towards the quasar J1439$+$1117: 
    spectrum (black curve) and simultaneous fit to the four transitions (red curve). 
   The bottom black curve in each panel corresponds to the low noise level (rms$\simeq$1.7\%). 
   At $v_{\rm helio}=0$~km~s$^{-1}$, the redshift is $z = 2.418372$: the corresponding 
   C\,{\sc i} component has the highest column density. 
  The blue tick marks in the C\,{\sc i}$\lambda$1656 panel (lower one) correspond to the 
   stronger seven components. In the C\,{\sc i}$\lambda$1560 panel, the two tick marks
  correspond to weaker components at $-$104 and $-$118 km~s$^{-1}$.
  In the upper panel, the absorption seen in the 50-90~km~s$^{-1}$ velocity
  range is a blend of C\,{\sc i}$^{\star}\lambda$1276 and C\,{\sc i}$\lambda$1277
  from the two weak components with the bluest velocity.
}       
   \label{J1439CI}      
   \end{figure}         

\subsubsection{ QSO J1439$+$1117: the C\,{\sc i} system at $z_{\rm abs}=2.41837$}

The UVES spectrum of this quasar has been discussed by Srianand et
al. (2008), who detected CO, H$_2$ , and HD molecules at $z_{abs} = 2.4184$.
Several C\,{\sc i} and C\,{\sc i}$^{\star}$ transitions (around 1276, 1277, and 1280 \AA) 
fall on the red part of the Ly$\alpha$ emission line while the stronger
multiplets around 1560 and 1656 \AA\ are seen against the continuum source
alone (unfortunately, transitions at $\lambda \approx 1328$ \AA\ fall in a
gap of the spectrum). We first examine whether a model in which the
absorber uniformly covers the whole source is consistent with all
absorption line profiles. When we attempted to fit the C\,{\sc i} and C\,{\sc i}$^{\star}$
transitions near 1276, 1277, 1280, 1560, and 1656 \AA, it clearly
appeared that additional absorption is present, blended with the 1277 multiplet (it
is presumably  Ly$\alpha$ at $z_{abs} = 2.59213$; when we adopt the emission
redshift $z_{em} = 2.5853 \pm 0.0001$ inferred from SDSS data, this
corresponds to a relative velocity of $V = - 570$ km s$^{-1}$). We therefore
retain only the 1276, 1280, 1560, and 1656 \AA\ multiplets. The simulateneous fit of 
these four transitions involves seven main velocity components (indicated by blue 
tick marks in the bottom panel of Fig. 11) that cover a range of 70 km s$^{-1}$. 
In addition, two weak detached components are also present at $v= -104$ and  $-$118 km 
s$^{-1}$ (these can be seen in the C\,{\sc i}~1560 panel and to a lesser extent, in the 
C\,{\sc i}~1280 panel); the corresponding C\,{\sc i}$\lambda$1277 features are blended with 
C\,{\sc i}$^{\star}$$\lambda$1276 absorption (upper panel of Fig. 11). 
\\
Some discrepancies between the observed spectrum and the fit are 
seen especially near C\,{\sc i}$^{\star}$ transitions (they are most apparent 
for the 1280 \AA\ multiplet), and to investigate the possibility of 
spatial variations over the ELR extent in more detail, we separately fit the 1560 and
1656 \AA\ multiplets (formed against the continuum source) and those at 1276, 1277, 
and 1280 \AA\ (formed against the ELR and continuum source; the 
Ly$\alpha$ absorption line mentioned above was included). 
Comparison of the two fits confirms that C\,{\sc i}$^{\star}$ absorption tends 
to be stronger on average towards  the ELR than towards 
the continuum source. 
The total C\,{\sc i}$^{\star}$ column density derived
from the 1276, 1277, and 1280 \AA\ multiplets, $N$(C\,{\sc i}$^{\star}$) $= 2.16 \pm
0.11 \times 10^{14}$ cm$^{-2}$ , is significantly higher than the value
derived from the 1560-1656 multiplets, $N$(C\,{\sc i}$^{\star}$) $= 1.50 \pm 0.04 \times
10^{14}$ cm$^{-2}$ (the corresponding values for C\,{\sc i} are nearly identical,
$N$(C\,{\sc i}) $\approx 4.1\times 10^{14}$ cm$^{-2}$). Since the two fits involve
velocity components with nearly identical redshifts, it is possible to compare the absorption toward the continuum 
and that towards the ELR separately. To this purpose, we used Eq. (7) to extract the absorption 
profile toward the ELR alone for the 1276 and 1280~\AA\ transitions
and adopted 
$x \simeq 3.1$ and 2.0, respectively, as measured on the spectrum.
\\
The result is shown in Fig. 12, where both the ELR absorption computed from Eq. (7) 
(red curve)  and the continuum source absorption derived from fitting the 1560 
and 1656~\AA\ transitions are shown (green curve). Absorption clearly tends to be weaker 
towards the continuum source, especially for C\,{\sc i}$^{\star}$. 
By fitting the 1276 and 
1280~\AA\ ELR profiles, we can compare the values of $N$(C\,{\sc i}) and 
$N$(C\,{\sc i}$^{\star})$ towards the continuum source and ELR for each 
component. For C\,{\sc i}, only the $v = -10$~km~s$^{-1}$ component shows a 
significant difference, with $N$(C\,{\sc i}) larger towards the ELR by 
a factor $2.1 \pm 0.15$. For C\,{\sc i}$^{\star}$, the three central components 
at $v = -10, 0$ and $+12$~km~s$^{-1}$ display a higher column density towards 
the ELR by factors of $3.0 \pm 0.3$, $1.4 \pm 0.1,$ and $2.1 \pm 0.2$, respectively. 
We conclude that this system shows significant spatial structure at scales 
in the range of 100~au - 0.1~pc. 

   \begin{figure}
\hspace{3mm}
        \includegraphics[width=9.5cm,angle=-90]{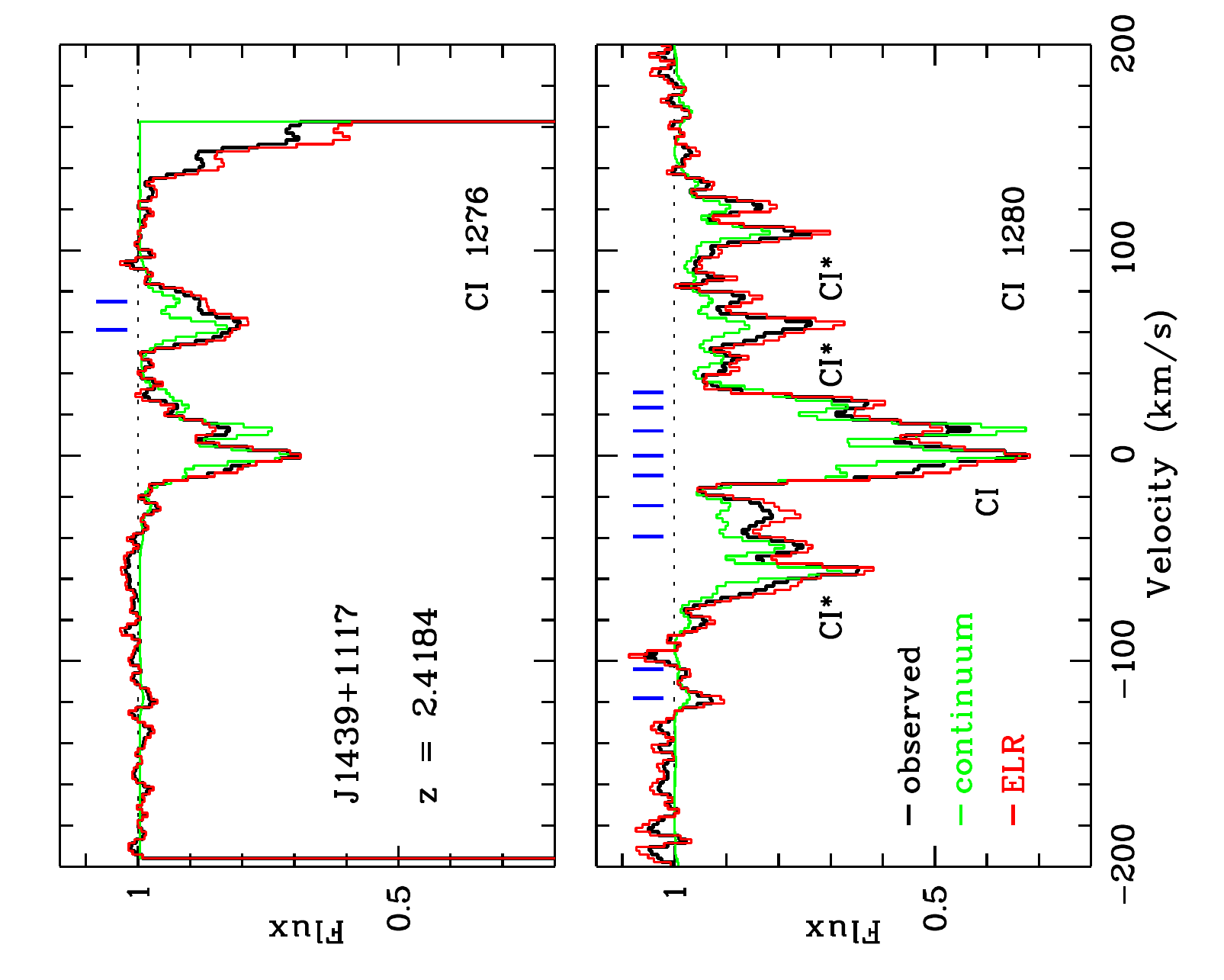}
    \caption{Normalized profile of the C\,{\sc i}, C\,{\sc i}$^{\star}$ , and 
C\,{\sc i}$^{\star \star}$ lines near 1276 (top panel) and 1280~\AA\ (bottom) 
at $z_{abs} \simeq 2.418$ towards J1439+1117, seen on the Ly$\alpha$ 
emission line of this quasar. Observed 
profiles are shown in black, while the profiles derived from fitting of 
the 1560 and 1656~\AA\ transitions observed against the continuum source are 
displayed in green and the profiles computed towards the ELR alone (see 
text) in red. C\,{\sc i}$^{\star}$ absorption is significantly stronger towards 
the ELR for the three central components (at $v = -10, 0$ and $+12$~km~s$^{-1}$) 
as well as the C\,{\sc i} absorption in the $v = -10$~km~s$^{-1}$ component. 
The velocity scale is the same as in Fig. 11. 
}       
   \label{JJ1439}       
   \end{figure} 


\subsubsection{PKS 1448$-$232: the Ca\,{\sc ii} system at $z_{\rm abs}=-0.00002$ }

The quasar PKS 1448$-$232 shows strong multicomponent absorption in  
Galactic Ca\,{\sc ii} and Na\,{\sc i} (Ben Bekhti et al. 2008). The Ca\,{\sc ii} 
absorption coincides with Ly$\alpha$ and N\,{\sc v} emission. The two doublet lines have 
distinct $x$ values (0.66 and 0.60 for the Ca\,{\sc ii}$\lambda$3934 and 3969 
lines, respectively),  
which offers the opportunity of investigating non-uniformity effects. Assuming that 
the Ly$\alpha$ ELR has an extent of about 1 pc, it delineates an angle $\theta
\approx 10^{-9}$ rad or 0.20 mas, 
given the quasar emission redshift, $z_{em} = 2.208$. This
corresponds to a very small linear extent of 0.02 au in a galactic cloud
located at a distance of about 100 pc. Since significant structure is
detected in Galactic Ca\,{\sc ii} gas only at scales on the order of 10 au
or higher (Smith et al. 2013; McEvoy et al. 2015), the absorber is expected
to be uniform and provides a test case for our fitting procedure. An
excellent fit is obtained for the whole Ca\,{\sc ii} doublet line profiles
with $C_f=1$, which confirms the absence of any departure from uniformity for
all pieces of the Galactic gas associated with the five main velocity components.

\subsubsection{ FBQS J2340$-$0053: the  C\,{\sc i} system at $z_{\rm abs}=2.05454$ }

   \begin{figure}
\hspace{-7mm}
        \includegraphics[width=10.cm,angle=0]{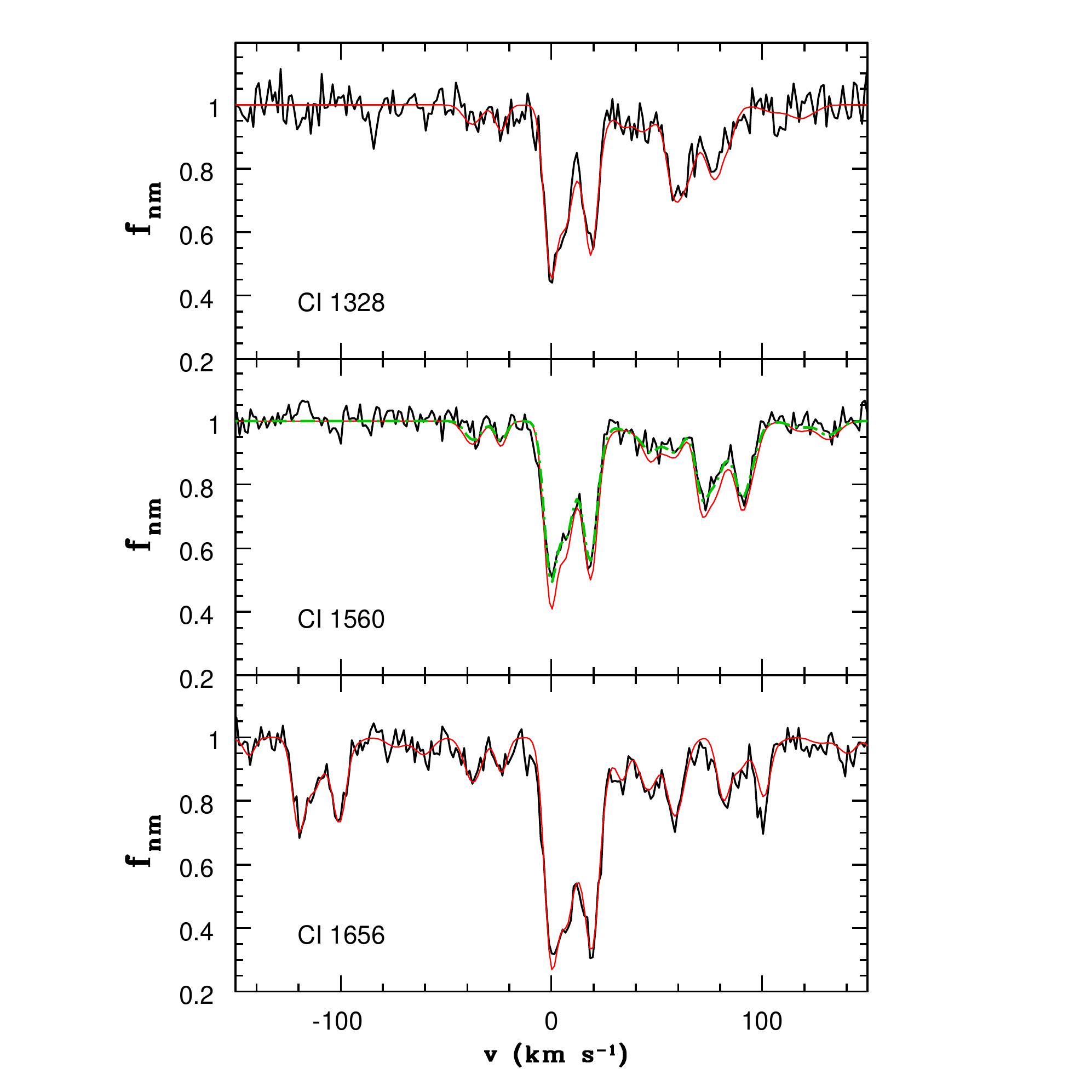}
    \caption{Intervening C\,{\sc i} absorption towards FBQS J2340$-$0053: 
    spectrum (black curve) and simultaneous fit obtained with the four transitions 
    on the quasar continuum (red curve). The green curve for the 
    C\,{\sc i}, C\,{\sc i}$^{\star}$$\lambda$1560 absorptions corresponds to a fit with 
    a spatial coverage factor $C_f=0.85$.  
   At $v_{\rm helio}=0$~km~s$^{-1}$, the redshift is $z = 2.054526$. }
   \label{J2340CI}      
   \end{figure} 

There are two available spectra taken  about two years apart (see Table 1). The C\,{\sc i}$\lambda$1560 
multiple absorption falls on the top of the C\,{\sc iv} quasar emission line; this region is only 
covered by the HIRES spectrum. A good fit of the  C\,{\sc i} lines that fall on the quasar continuum 
(C\,{\sc i}$\lambda$1277,1280,1328 just redwards of the Ly$\alpha$-N\,{\sc v} 
quasar emission line, and C\,{\sc i}$\lambda$1656) is obtained separately for the HIRES and UVES 
spectra using eight components and assuming full coverage of the quasar continuum, $C_{c} = 1$. 
The estimated values of $N$ and $b$ are consistent between the two epochs, except for the C\,{\sc i} 
component at $z_{\rm abs}=2.05472,$ for which tentative variation has been reported
by Boiss\'e et al. (2015). 
\\
The physical properties (density and temperature) of this  multiple-component absorption system 
have been thoroughly investigated by Jorgenson et al. (2010) using the C\,{\sc i} fine-structure 
lines and H$_2$ absorption. Their  C\,{\sc i} fit is solely based on the HIRES data; it includes 
an unidentified line blended with C\,{\sc i}$\lambda$1328 (Jorgenson: private communication) 
and does not consider spatial covering effects of the C\,{\sc iv} ELR.   
\\
Our fit of the  HIRES data obtained for the four transitions that fall on the quasar continuum 
is shown in Fig. 13 for two transitions (C\,{\sc i}$\lambda$1328 and 1656) as well as for 
C\,{\sc i}$\lambda$1560, which is on the C\,{\sc iv} quasar emission line. 
The C\,{\sc i}$\lambda$1560 and C\,{\sc i}$^{\star}$$\lambda$1560 multiple-component profiles  
are both overfitted (Fig. 13 red curve), which points towards 
partial covering of the  C\,{\sc iv} emission line region. We then applied a 
spatial coverage factor to these absorption lines: the best fit is obtained for $C_f=0.85$ 
(with an uncertainty of $\pm0.05$), which is also plotted in Fig. 13 (green curve). 
\\
The determination of the coverage factor of the emission line region $C_{elr}$ requires an 
estimate of the quasar continuum flux $F_c$ that underlies the C\,{\sc iv} emission line. 
There is a 2004 SDSS low-resolution spectrum of this quasar 
that we use to derive a flux ratio $x = 0.68$. 
This yields $C_{\rm elr}=0.63$ with a possible range of $0.50<C_{elr}<0.75$. 

\subsection{Absorption systems with H$_2$ lines on Ly$\beta$-O\,{\sc vi} emission}

$\bullet$  FBQS J2340$-$0053, the H$_2$ absorber at $z_{\rm abs}=2.05473$\\
In this system, numerous transitions from H$_2$ are seen from levels $J=0$ up to $J=5$
(Jorgenson et al. 2010). We examined the spectrum in detail in the range containing
the quasar Ly$\beta$-O\,{\sc vi} emission line and found several features from 
$J=0$ (at 1049.37~\AA), $J=1$ (at 1049.96~\AA), $J=2$ 
(1040.37~\AA), and $J=3$ (1041.16 and 1043.51~\AA) that reach the zero 
level at the core of absorption lines associated with the main velocity components
(we estimate that the residual flux is lower than about 3\% at
the 3$\sigma$ level). This unambiguously indicates that the 
Ly$\beta$-O\,{\sc vi} ELR is fully covered ($C_{elr} \simeq 1$) by both the cold molecular 
material ($J=0$ and $J=1$) and the higher excitation gas ($J \geq 3$) in these 
velocity components.

Other intervening systems have been studied in the literature for which some constraints 
can be derived. To our knowledge, the only marked partial covering effect of an ELR by 
an H$_2$ absorber has been found at $z_{\rm abs}=2.338$ towards 
LBQS 1232$+$082 (Ivanchik et al. 2010; Balashev et al. 2011). The covering factor of the 
O\,{\sc vi} ELR is $C_elr = 0.8$ (see Table 2 and Fig. 4 in Balashev et al. 2011). 
\\
For the system at $z_{abs} = 2.40183$ towards the quasar HE~0027$-$1836, 
Noterdaeme et al. (2007) did not discuss potential partial covering effects but presented a set 
of fits for H$_2$ line profiles from levels $J=0$ up to $J=5$, including lines seen 
either on or off of the Ly$\beta$-O\,{\sc vi} quasar emission line (given the quasar redshift, 
$z_{em}=2.55$, H$_2$ transitions coinciding with this emission have 
$1075 \leq \lambda \leq 1091$~\AA). This is a single-component system with low 
H$_2$ column density  ($N$(H$_2$) $\simeq 2.\times 10^{17}$~cm$^{-2}$), and 
the strongest $J=0$ or $J=1$ lines are not flat-bottom and barely reach the zero level. 
However, no mismatch is observed in the fits for lines occurring on 
Ly$\beta$-O\,{\sc vi} emission, therefore we can rule out marked partial 
covering or structure effects. 
\\
In the $z_{\rm abs}=2.059$ system towards QSO J2123$-$0050,  
Klimenko et al. (2016) found a 3\% residual flux at the core of high-opacity $J=0$ 
and $J=1$ H$_2$ lines seen on Ly$\beta$-O\,{\sc vi} emission, implying $C_{elr}$ is close 
to but significantly lower than 1. 
\\
Finally, for the $z_{\rm abs}=2.811$ system towards PKS 0528$-$250 
Klimenko et al. (2015) established the reality of a 2\% residual flux; 
however, this blazar is nearly devoid of emission lines and the residual is seen regardless of 
the location of absorption lines in the spectrum. The interpretation favoured 
by these authors  is that the continuum source itself is extended and includes a 
jet that is not entirely covered by the absorber. 
        
\section{Estimate of the transverse absorber extent}

In the following, we discuss the implications of the results presented above about the 
extent and structure of the various types of the foreground absorbers we considered, involving 
either neutral, diffuse molecular, or moderately ionised gas. 
Since partial covering or 
structure effects involve the relative projected size of the absorber and ELR, we first 
need to review which information we have about the extent of the latter. 

\subsection{Estimate of the ELR size}
The size-luminosity scaling relation provided by C\,{\sc iv} reverberation mapping studies 
of high-redshift luminous quasars (Trevese et al. 2014, and references therein) can be used 
to estimate  the ELR size, $l_{elr}$. 
For instance, applying this relation to HE~0001$-$2340 yields an ELR overall size 
of about 0.3 pc. This value should be appropriate for other quasars as well since 
all targets in our sample have about the same luminosity and redshift, and since its 
dependence on luminosity $L$ is moderate (roughly $\propto L^{0.5}$). We note that 
some fraction of Ly$\alpha$ emission can extend beyond the C\,{\sc iv} ELR 
(Balashev et al. 2011; Fathivavsari et al. 2016; Fathivavsari et al. 2017). 
\\
In the following, when we discuss the constraints derived on the absorber extent, we assume for simplicity that the ELR displays a uniform brightness with the same extent at 
all  relative velocities. 

\subsection{Extent and structure of neutral absorbers} 
\subsubsection{Unambiguous detections of non-uniform absorbers}

We outline the three cases with marked structure effects in neutral 
gas traced either by Fe\,{\sc i} or C\,{\sc i}. 
\\
$\bullet$ HE~0001$-$2340, the Fe\,{\sc i} absorber at $z_{\rm abs}=0.45206$ 
\\
This is our only clear case of small ($<50$\%) spatial covering of the ELR,
corresponding to the configuration shown in Fig. 1a. The analysis of line 
profiles and the curve-of-growth method both yield a spatial coverage factor $C_f=0.32$ 
of the Ly$\alpha$ ELR (see Sect. 3.2.2).
This value is fully compatible with $C_f$ being at its minimum, $C_{f,min}=0.30$ 
(as derived from the measured flux ratio $x$; see Figs. 5 and 6), that is, with a 
spatial coverage of the ELR $C_{elr} \simeq 0$, together with full coverage of 
the continuum source.
An estimate of the maximum value allowed by the data is $C_f \simeq 0.37$,
implying $C_{elr,max}\simeq 0.10$. 
The overall extent, $l_{abs}$, of the absorber must then be quite small. When we 
take the different redshifts of the absorber and quasar 
into account (0.3~pc at $z_{em} = 2.28$ defines the same angle as 0.21~pc at $z_{abs} = 0.45$),  
$C_{elr,max}$ scales as $(l_{abs}/l_{elr})^2$, we obtain a maximum absorber overall size 
$l_{abs} \simeq 0.06$~pc, which is well consistent with the range 0.01 - 0.6pc inferred 
from modelling by Jones et al. (2010).
\\
The spectroscopic characteristics  of this absorber are similar to those of the so-called CaFe 
clouds (Bondar et al. 2007). The estimate of their physical properties depends on the assumed 
depletion level: either insignificant, which results in a high gas temperature $T\sim 
5000$-10000 K (Gnaci\'nski  \& Krogulec 2008), or strong, with an assumed high H$_2$ fraction 
and low temperature $T < 100$~K (Jones et al. 2010). 
We did not detect any absorption from CH and CH$^+$, although this does not exclude the presence of 
dust and moderate depletion, as cautioned by Welty et al. (2008).  
\\
$\bullet$  QSO J1439$+1117$, the  C\,{\sc i} absorber at $z_{\rm abs}=2.41837$ 
\\
For this system, one C\,{\sc i} and three C\,{\sc i}$^{\star}$ components are found to display 
higher column density values towards the ELR than to the continuum source, with  
ratios ranging from 1.4 to 3.0. This means that significant spatial structure must be present at 
scales in the 100~au - 0.1~pc range, with the continuum source seen through more tenuous and less 
dense gas at the velocities of these three components. However, since the strongest components nearly 
reach the zero level in C\,{\sc i}$\lambda$1277, the associated gaseous clouds 
should cover the whole ELR.\\

$\bullet$  FBQS J2340$-$0053, the C\,{\sc i} absorber at $z_{\rm abs}=2.05473$ 
\\
This  C\,{\sc i} multiple absorption system has five well-detected transitions, one of 
which is on the C\,{\sc iv} quasar emission line. The constraints obtained from 
the analysis of line profiles imply partial covering  of the ELR with  
$C_{elr}=0.63$ for the C\,{\sc i} and C\,{\sc i}$^{\star}$ components (see Fig. 13). 
This case corresponds to the configuration shown in Fig. 1b.  The minimum size of 
the absorber, $l_{abs,min}$, is obtained when the whole absorber covers the ELR,  
for which we obtain $l_{abs,min}=0.79\,l_{elr}$. 
All values above $l_{abs,min}$ are possible for $l_{abs}$. This direct constraint 
on the transverse size can be compared to the indirect estimate derived by 
Jorgenson et al. (2010) from an analysis of the physical properties of the gas
(density in particular). For each main C\,{\sc i} component (velocity range 0 - 25 
km s$^{-1}$) these authors give an extent along the line of sight of about 1.5 pc, 
which is quite consistent with our results. For such large sizes, 
the gas associated with the $v\simeq 19$~km s$^{-1}$ component is not expected to display time 
variability if it were uniform. Then, if the tentative variations detected by Boiss\'e 
et al. (2015) between the 2006 and 2008 spectra are real, structure in the 
10 - 100~au range must be present (we note that the velocity scale adopted in 
Fig. 3 from Boiss\'e et al. is different from that in this paper). 

\subsubsection{Possible detections of non-uniformity effects }

We summarize the three cases with possible spatial covering effects for neutral 
gas traced either by Fe\,{\sc i} or C\,{\sc i}.  \\
$\bullet$ PKS 0237$-$23, the C\,{\sc i} absorber at $z_{\rm abs}=1.36469$ 
\\
This  C\,{\sc i} multiple component absorption system has only two detected transitions: 
one on Ly$\alpha$ emission, and the other in the  Ly$\alpha$ forest. 
The narrowest component at $z_{\rm abs}=1.36469$ is partly resolved  
($b=4.3$ km s$^{-1}$). 
The line profile analysis with partial covering of the ELR implies 
 $C_f=0.85$ and $C_{elr}=0.78$, thus a size $l_{abs}$ of about 0.23 pc.\\
An alternative model involves a 'hidden' very narrow component together with full spatial 
covering of the ELR. The line profile analysis for this narrow component yields a width 
$b=0.5$ km s$^{-1}$; this model is not favoured considering the fairly 
high temperature ($T \gtrsim 1000$ K) inferred for the C\,{\sc i} phase. 
\\
$\bullet$ TXS 1331$+$170, the Fe\,{\sc i} absorber at $z_{\rm abs}=0.74461$ 
\\
This Fe\,{\sc i} single-component absorption system has five well-detected transitions, 
one of which is on the Ly$\alpha$-N\,{\sc v} quasar emission line. The latter is weak and
somewhat overfit with full spatial covering. A better fit is obtained with $C_f\sim 0.8$, 
thus $C_{elr}=0.56$ for $x=0.83$. However, this is not highly significant considering 
the small differences in the fits relative to the rms of the spectrum.  
\\
$\bullet$ TXS 1331$+$170, the C\,{\sc i} absorber at $z_{\rm abs}=1.77653$ \\
Three well-detected transitions are present, one of which is on the 
blue wing of Ly$\alpha$ emission: we obtain  $C_f\gtrsim 0.9,$ which for $x=0.41$ yields 
$C_{elr} \gtrsim 0.66$; however, $C_f=1.0$ is fully acceptable. 
Consequently, this cold C\,{\sc i} absorber does not 
show an unambiguous spatial covering effect, which is consistent with the presence of 
associated H$_2$ saturated absorptions with flat cores and zero residual intensities.  

\subsubsection{Statistical analysis}

We now attempt to extract quantitative information on the neutral absorber extent from results 
obtained for the whole sample. Clearly, the population is not homogeneous since the Fe\,{\sc i} 
absorber at $z_{abs}=0.452$ towards HE~0001$-$2340 covers a negligible 
fraction of the ELR, while the others cover all or most of it. We therefore exclude the former 
from our overall analysis. Although the remaining six systems provide poor statistics (we 
include the Ca\,{\sc ii} system towards HE~0001$-$2340 at $z_{abs}=0.270$ 
since Ca\,{\sc ii} traces regions containing gas that is mainly neutral), the fact that only 
one of them shows unambiguous non-uniformity effects tells us that generally, when these absorbers 
cover the continuum source, they also cover most of the ELR. Their extent must then be 
notably larger than that of the ELR. In the following, we therefore consider that of the six 
absorbers that cover the continuum source, at least one displays partial covering with 
$C_{elr} \leq 0.75$, which leads to a probability $P(C_{elr} \leq 0.75) \geq 0.17$. \\
To analyse the implication of this rough estimate in terms of relative size, 
we assume for simplicity i) that all these six absorbers display a spherical shape
with the same radius, $R_a$, and ii) that the effect of ELR and absorber redshift 
differences can be ignored (we proceed as if the ELR and absorber had the same redshift). 
In projection, the ELR is seen as a disc 
(radius $R_e$), and the covering factor, $C_{elr}$, is simply given by the ratio 
$C_{elr}=A/(\pi R_e^2)$, where $A$ is the area of the region over which the absorber
and ELR discs overlap. Let $d$ be the projected distance between the 
centre of the ELR and the absorber discs. 
Following Ofengeim et al. (2015), $A$ can be expressed as 
\begin{equation}
A = R_a^2\arccos (y_a) + R_e^2\arccos (y_e)-H,
\end{equation}
with 
\begin{equation}
y_a=\frac{d^2+R_a^2-R_e^2}{2dR_a},\,\,\, y_e= \frac{d^2+R_e^2-R_a^2}{2dR_e},
\end{equation}
and
\begin{equation}
H=\frac{\sqrt{((R_e+R_a)^2-d^2)(d^2 - (R_a-R_e)^2)}}{2}.
\end{equation}
Consider for instance a model in which $R_a = R_e$ and compute the expected probability 
of obtaining $P(C_{elr}\leq 0.75)$. Of the absorbers that cover the continuum source (i.e., those that
satisfy $d<R_a$), Eq. (14) implies that only those with $d \geq 0.40\,R_e$ have 
$C_{elr}\leq 0.75$. Then, for $R_a = R_e$,
\begin{equation}
P(C_{elr} \leq 0.75) = \frac{\pi R_a^2 -\pi (0.40\,R_e)^2}{\pi R_a^2} = 0.84. 
\end{equation}
The latter 
value would imply too many absorbers with marked partial covering effects as compared to 
what we observe. Proceeding in the same way for higher $R_a/R_e$ values, we find that 
$P(C_{elr}\leq 0.75) = 0.17$, for $R_a\simeq 5\,R_e$. The latter value is to be considered 
as an upper limit for $R_a$ since some of the systems noted ``possible'' might show 
real partial covering.
We conclude that a neutral absorber radius of about $5\times 0.15= 0.75$~pc or smaller 
is consistent with the data. Interestingly, these values are in the range (0.2 - 4.7~pc) derived by 
Jenkins \& Tripp (2011) for Galactic C\,{\sc i} clouds.

\subsubsection{Diffuse molecular gas}

C\,{\sc i} is known to be nearly cospatial to H$_2$ (Srianand et al. 2005),
and it is therefore relevant to compare
the extent derived for these two species. As discussed above (Sect. 3.3), constraints are 
available for very few H$_2$ absorbers. For the only case in which a marked partial covering 
effect is found, the absorber towards LBQS 1232$+$082, the estimated 
$C_{elr}$ value ($\simeq 0.7$) is comparable to those obtained for C\,{\sc i} absorbers, 
leading to a lower limit of a few 0.1~pc size. \\
It is noticeable 
that for the FBQS J2340$-$0053 absorber, the excited H$_2$ ($J=3$) lines display no residual 
flux. This provides useful constraints for models invoking intermittent turbulent vortices 
to account for this warmer gas (Godard et al. 2014): although these localized regions are 
supposed to fill a tiny volumic fraction, their number density and extent must be such that 
their surface coverage factor is very close to unity.\\

\subsection{Singly-ionised gas}

When possible, we searched for structure effects in moderately ionised 
gas, as traced by species such as Mg\,{\sc ii}, Fe\,{\sc ii,} and Mn\,{\sc ii}. 
\\
The only conclusive detection of partial covering of the ELR for a singly-ionised 
species is that of Fe\,{\sc ii} at $z_{\rm abs}=0.72609$ towards Tol 0453$-$423. 
The effect is small, but very unusual: the Fe\,{\sc ii}$\lambda2600$ saturated  
components have a flat core with a clear flux residual of 2\% (7$\sigma$ 
significance level). We obtain from the fit of this Fe\,{\sc ii} doublet, which 
falls on the quasar Ly$\alpha$-N\,{\sc v} emission, $C_f=0.98$ and $C_{elr}=0.96$, 
that is, an absorber size comparable to or larger than the ELR. 
\\
For two cases, either full or partial covering of the ELR are 
equally acceptable or inconclusive due to the weakness of the absorption lines:
the Mg\,{\sc ii} absorber at $z_{\rm abs}=0.452$ towards HE 0001$-$2340, 
with $0.75 \lesssim C_f \leq 1.0$, and the very weak Mn\,{\sc ii} absorber 
at $z_{\rm abs}=0.726$ towards Tol 0453$-$0423. 
\\
Finally, two cases  are well consistent with full covering of the ELR: 
the multiple Fe\,{\sc ii} system at $z_{\rm abs}=1.365$ towards PKS 0237$-$23 
and the multiple Fe\,{\sc ii} system at $z_{\rm abs}=1.328$ towards  
TXS 1331$+$170; in total 22 distinct velocity components are detected in these Fe\,{\sc ii} 
absorbers. 
The absence of structure effects in these Fe\,{\sc ii} absorbers appears to be in good 
agreement with constraints obtained previously concerning the extent - about 200 pc 
after Rauch et al. (2002) - of low-ionisation cloudlets associated with individual 
velocity components. 

\section{Discussion and prospects}

\subsection{Detection of non-uniformity effects}  

The analysis of the systems in our sample illustrates well the difficulties 
encountered in establishing the reality of non-uniformity effects in foreground absorbers. 
Large departures from uniformity as seen towards HE~0001$-$2340 for 
Fe\,{\sc i} at $z_{abs}=0.45206$ are very rare, and the observed effects are generally too small 
to bring equivalent width or apparent opacity ratios clearly out of the range expected for a 
point-like background source.  
When the values of these ratios remain within the optically thin and thick bounds, the apparent 
inconsistencies between the various line profiles can potentially be removed by introducing ``hidden'' 
additional velocity components (see the $z_{abs}=1.364$ absorber towards
 PKS~0237$-$23: we note, however, that for this system, additional arguments do not 
favour the presence of a low $b$ component). In the interstellar medium of 
our own Galaxy, ultra-high resolution observations have 
directly shown that $b$ parameters in some neutral velocity components can be lower than
$b=1$~km s$^{-1}$ (Andersson et al. 2002); our line fitting or curve-of-growth analysis also 
require similarly narrow lines, but for severe blending, our resolution 
of about 6~km s$^{-1}$ might be insufficient to reach reliable conclusions. 

It is noteworthy, however, that when two transitions with about equal $\lambda\,f$ values 
(e.g., to within 10\%) are detected, one against the continuum and the other one on an 
emission line, robust results can be obtained; indeed, the opacities being then nearly 
equal, line profiles are expected to be identical in the velocity scale (corresponding to 
$W \propto \lambda$),  regardless of any assumption on the velocity distribution 
(Fe\,{\sc i}$\lambda 3021$ and Fe\,{\sc i}$\lambda 2719$ seen towards 
HE~0001$-$2340 nearly fulfill this condition), which results in a very narrow range expected 
for the $W$ ratio (Fig. 6). 
For the main C\,{\sc i} and Fe\,{\sc i} transitions, we note that the $\lambda\,f$ ratio of 
the C\,{\sc i}$\lambda 1277$ to C\,{\sc i}$\lambda 1328$ lines is 1.075,
 while the $\lambda\,f$ ratios of the Fe\,{\sc i}$\lambda 2719$ to Fe\,{\sc i}$\lambda 2167$ 
lines and the Fe\,{\sc i}$\lambda 2719$ to Fe\,{\sc i}$\lambda 3021$ lines 
are 1.018 and 1.054, respectively, which is very close to unity. 
This means that for a C\,{\sc i} or Fe\,{\sc i} system that displays one of these transitions 
on an emission line, it would be easier to investigate structure effects.

One obviously important parameter in detecting structure effects is the emission line 
over continuum contrast, that is  the $x$ value. If the latter is high 
(as with absorption detected on top of Ly$\alpha$ emission), small departures from full 
covering can be revealed more easily. In practise, for
the high $z$ quasars considered in this study, Ly$\alpha$-N\,{\sc v} or C\,{\sc iv} 
emission, and to a lesser extent Ly$\beta$-O\,{\sc vi}, are the only emission lines 
providing reasonably high $x$ values. Another key parameter is the absorption line 
opacity: when optically thick flat-bottom features are seen on emission lines, 
departures from full covering of only a few percent can be established reliably, as 
illustrated by the $z_{abs}=0.726$ system towards Tol 0453$-$423. For 
low- or intermediate-opacity lines, the detection of several transitions with various $f$ values 
formed against either the continuum source or ELR is important to achieve well-constrained 
fits for absorption line profiles and to assess the uniformity of the absorber. 
Species such as Fe\,{\sc i} and C\,{\sc i} are well suited owing to the large number of 
relatively strong transitions they provide.

Finally, we pinpoint a few values of the $r =(1+z_{em})/(1+z_{abs})$ ratio that provide 
especially favourable distributions of Fe\,{\sc i} and C\,{\sc i} 
absorption lines with respect to quasar emission lines. We first consider the main C\,{\sc i} 
transitions and we find that $r \simeq 1.072$ brings the 1260, 1276, 1277, and 1280~\AA\ 
transitions on Ly$\alpha$-N\,{\sc v} together with C\,{\sc i}$\lambda$1656 on C\,{\sc iv} 
emission (this corresponds to the system observed by Balashev et al. 2011 with $z_{em}=2.57$ 
and $z_{abs}=2.338$, thus $r =1.070$). For $r \simeq 1.035$, the 1260, 1276, 1277, and 
1280~\AA\ transitions lie on Ly$\alpha$-N\,{\sc v,} while all other transitions are seen 
against the continuum source alone. For Fe\,{\sc i}, we find that $r \simeq$ 2.512 
(Fe\,{\sc i}$\lambda$3021 on Ly$\alpha$ and Fe\,{\sc i}$\lambda$3720 on C\,{\sc iv}),
2.239 (Fe\,{\sc i}$\lambda$2719 on Ly$\alpha$ and Fe\,{\sc i}$\lambda$3441 on 
C\,{\sc iv}, as towards HE~0001$-$2340) and 1.928 (Fe\,{\sc i}$\lambda$3021 
on C\,{\sc iv} and Fe\,{\sc i}$\lambda$3720 on C\,{\sc iii}]) provide valuable configurations.

\subsection{Incomplete absorption systems} 

The discussion in Sect. 4.2 and our conclusion that the extent of cold 
neutral absorbers is about five times that of the ELR raises the possibility
of detecting absorption lines induced by cloudlets that  cover part of the 
ELR, but not the continuum source (Fig. 1c). This would correspond to velocity 
components seen only on emission lines, with no corresponding feature detected in
transitions seen against the continuum.  Such small clouds are expected  
to be embedded in a more extended HI region 
 and in an even larger ionised gaseous envelope. Thus, one does 
not expect to miss the identification of systems that would present
absorption lines exclusively on emission lines, but instead to occasionally 
encounter some additional velocity components on the emission line.

The simple model described in Sect. 4.2.3 can be used to estimate the incidence of such 
components. The associated absorbers are characterised by 
$R_a < d < R_a + R_e$; then, their number ($N_e$) relative to the number of those 
covering the continuum source ($N_c$) is given by 
\begin{equation}
\frac{N_e}{N_c} = \frac{\pi (R_a+R_e)^2-\pi R_a^2}{\pi R_a^2} =
\left(1+\frac{R_e}{R_a}\right)^2-1.
\end{equation}
For $R_a \simeq 5\ R_e$, we obtain $N_e/N_c = 0.44$. A significant 
fraction of incomplete systems is therefore expected, but the latter 
should on average exhibit much weaker features since the absorbed flux is 
low. $C_c=0,$ and furthermore, for $R_a = 5\ R_e$, we find 
that 50\% of these systems have $0<C_{elr}<0.17$ (with $d$ lying in the range  
$5.52\ R_e < d < 6\ R_e$); the remaining 50\% have $0.17<C_{elr}<0.48$ and 
$5\ R_e < d < 5.52\ R_e$. We searched for such additional components 
within the C\,{\sc i} and Fe\,{\sc i} systems investigated in this paper  
and could not find any. To our knowledge, no absorption feature of this type 
has been reported. 

\subsection{Structure effects and $f$ values}

Some oscillator strength values for UV transitions that are commonly detected in quasar absorption systems 
are still not accurately determined. In particular, there has been some debate on the $f$ values for 
C\,{\sc i} transitions as derived by Jenkins \& Tripp (2001 \& 2011) from interstellar absorption 
lines using 
high spectral resolution HST-STIS data, and those given by  Morton (2003). When precise laboratory 
measurements or computations are not available, astrophysical data can be used to constrain the ratio 
of $f$ values for several transitions from a given species (see, e.g., Federman \& Zsargo 2001).  
 The excellent S/N quasar spectra considered in this paper could help to determine which 
of the two sets of C\,{\sc i} $f$ values is the most consistent with the observed line profiles. 

In doing this, transitions that occur on emission lines must of course be dismissed 
because a change in $f$ and partial covering of the ELR might be confused.  Equation (2) indicates 
that for low opacities, $F_{obs,n}$ reduces to  $F_{obs,n} \simeq 1-\tau \, C_f $ where 
$\tau \propto f$; thus both effects are indistinguishable in the $\tau \ll 1$ limit. 
All our measurements are consistent with full covering of the  continuum source, as 
indeed expected given the small size of the latter. Then, for 
transitions seen against the continuum, quasar spectra can provide reliable constraints on the 
ratio of $f$ values, exactly as stellar spectra do for interstellar lines. 

We therefore selected pairs of C\,{\sc i} transitions seen on the quasar continuum, 
when possible in regions of similar S/N, and with Morton's $f$ ratios differing substantially from 
those of Jenkins  \& Tripp.  This is clear for the 
f(C\,{\sc i}$\lambda 1560$)/f(C\,{\sc i}$\lambda 1656$) ratio: using Jenkins \& Tripp, it is 
1.74 times that obtained with Morton's $f$ values (there is a similar difference for 
C\,{\sc i}$^{\star}$ transitions). It is also the case, but to a lesser extent, for the 
f(C\,{\sc i}$\lambda 1560$)/f(C\,{\sc i}$\lambda 1277$) ratio, which is 1.13 times 
higher when adopting the values of Jenkins \& Tripp instead of
those of Morton.
\\
Two systems with C\,{\sc i}$\lambda 1560$ and C\,{\sc i}$\lambda 1656$ transitions are detected 
against the continuum: at $z_{\rm abs}=1.77653$ towards TXS~1331$+$170 
(HIRES spectrum), and at $z_{\rm abs}=2.41837$ towards QSO~J1439$+$1117 
(UVES spectrum). 
The simultaneous fit of these lines in both cases results in a much poorer fit when using the 
$f$ values of Jenkins \& Tripp instead of those of Morton. For TXS~1331$+$170, the $\chi^2$ 
statistic for the fit is 33\% higher, with the C\,{\sc i}$\lambda 1560$ transition of the 
moderate-opacity component at $-17$ km s$^{-1}$ (see Fig. 10) clearly overfit, whereas 
it is underfit for the C\,{\sc i}$\lambda 1656$ transition. 
The system at $z_{\rm abs}=2.41837$ toward QSO~J1439$+$1117 has multiple components with strong 
C\,{\sc i}$^{\star}$ absorptions of intermediate opacity. The $\chi^2$ statistics for the fit 
is 95\% higher with Jenkins \& Tripp's $f$ values, corresponding to a clear overfit of
the C\,{\sc i}$^{\star}$$\lambda 1560$ transitions and significant underfit for the 
C\,{\sc i}$^{\star}$$\lambda 1656$ lines. In contrast, satisfactory fits are obtained
for both systems using Morton's values.
\\
The C\,{\sc i}$\lambda 1560$ and C\,{\sc i}$\lambda 1277$ transitions are present at 
$z_{\rm abs}=1.77653$ in the UVES spectrum of TXS~1331$+$170. Since the $f$ ratios for 
these transitions only differ by 13\% between Jenkins \& Tripp's and Morton's values, we do 
not expect a strong difference in the fits. 
The $\chi^2$ statistics for the fit is only 9\% higher with Jenkins \& Tripp's $f$ values 
instead of Morton's values.  In both cases, the fit of C\,{\sc i}$\lambda 1560$ absorptions is 
good (the S/N ratio is better near the C\,{\sc i}$\lambda 1560$ lines, which drive the fit), 
whereas the C\,{\sc i}$\lambda 1277$ absorption of the component at $-17$ km s$^{-1}$ 
(see above) is slightly underfitted when using Jenkins \& Tripp's $f$ values. 
\\
Therefore, the $f$ values adopted for C\,{\sc i} in our study (from Morton 2003) appear to 
be more consistent with the data for the three systems mentioned above. 

Another species with transitions for which $f$ values are 
still uncertain is Ni\,{\sc ii} (see Cassidy et al. 2016 for a comparison between 
 calculations, experimental determinations, and observations). In the course of our 
study, we identified a few systems that we believe to be free of partial covering or 
structure effects; we plan to investigate them in a future paper in order to improve the 
determination of $f$ values for this ion.

\subsection{Prospects}

The conclusions drawn from our analysis are limited by the small number of systems we investigated, 
and it is clearly desirable to extend the C\,{\sc i} and Fe\,{\sc i} samples. Additional C\,{\sc i} 
absorbers can be selected with an appropriate $(1+z_{abs})/(1+z_{em})$ ratio from the C\,{\sc i} 
survey performed by Ledoux et al. (2015) on the basis of SDSS-II DR9 spectroscopic data. This 
could be complemented by a similar C\,{\sc i} search in the public SDSS-III DR12 spectroscopic 
database. 
For Fe\,{\sc i}, systems involving this species are much rarer and weaker than C\,{\sc i} 
systems, and it would be very difficult to build a sample of reasonable size. 
We note that some multi-object spectrographs 
such as MOONS at the VLT and 4MOST at VISTA, which are still under construction, will help in the 
near future to increase the number of high-redshift quasar identifications and will provide
additional C\,{\sc i} and Fe\,{\sc i} systems of interest for studies of cold neutral absorber 
extent and structure. 
   

Another promising approach would consist of using observations
with higher spectral resolution.
As illustrated by our study, at a resolution of about 6~km/s, some velocity components in neutral 
gas remain unresolved, and furthermore, line blending of adjacent components is often a problem in
establishing the reality of structure effects. The upcoming ESO instrument ESPRESSO, with 
its higher resolution (R=120000 and up to 220000) and full wavelength coverage from 3800 to 
7800 \AA\ (Pepe et al. 2013), 
would clearly allow us to perform a much more detailed study of the few systems for which
we have identified partial covering or non-uniformity effects. This is 
especially true for the complex system at $z_{\rm abs}=2.41837$ towards 
QSO~J1439$+$1117, which involves many closely spaced velocity components.

\begin{acknowledgements} 
We are grateful to the anonymous referee for the careful reading of the 
manuscript and numerous constructive comments. We also warmly thank Steve 
Federman, Pasquier Noterdaeme, and Serj Balashev for fruitful discussions. 
\end{acknowledgements}

\end{document}